# Electronic Spectroscopy of Isolated DNA Polyanions

Steven Daly,[1] Massimiliano Porrini,[1] Frédéric Rosu,[2,*] and Valérie Gabelica[1,*]

[1] Laboratoire Acides Nucléiques: Régulations Naturelle et Artificielle, Université de Bordeaux, Inserm & CNRS (ARNA, U1212, UMR5320), IECB, 2 rue Robert Escarpit, 33607 Pessac, France.

[2] Institut Européen de Chimie et Biologie, Université de Bordeaux, CNRS & Inserm (IECB, UMS3033, US001), 2 rue Robert Escarpit, 33607 Pessac, France.

* Frédéric Rosu (f.rosu@iecb.u-bordeaux.fr) or Valérie Gabelica, v.gabelica@iecb.u-bordeaux.fr

**Abstract:** In solution, UV-vis spectroscopy is often used to investigate structural changes in biomolecules (i.e., nucleic acids), owing to changes in the environment of their chromophores (i.e., the nucleobases). Here we address whether action spectroscopy could achieve the same for gas-phase ions, while taking the advantage of additional spectrometric separation of complex mixtures. We therefore systematically studied the action spectroscopy of homo-base 6-mer DNA strands ($dG_6$, $dA_6$, $dC_6$, $dT_6$) and discuss the results in light of gas-phase structures validated by ion mobility spectrometry and infrared ion spectroscopy, of electron binding energies measured by photoelectron spectroscopy, and of calculated electronic photo-absorption spectra. When UV photons interact with oligonucleotide polyanions, two main actions may take place: (1) fragmentation and (2) electron detachment. The action spectra reconstructed from fragmentation follow the absorption spectra well, and result from multiple cycles of absorption and internal conversion. The action spectra reconstructed from the electron photodetachment (ePD) efficiency reveal interesting phenomena: ePD depends on the charge state because it depends on electron binding energies. We illustrate with the G-quadruplex $[dTG_4T]_4$ that the ePD action spectrum shifts with the charge state, pointing to possible caveats when comparing the spectra of systems having different charge densities to deduce structural parameters. Moreover, ePD is particularly efficient for purines but not pyrimidines. ePD thus reflects not only absorption, but also particular relaxation pathways of the electronic excited states. As these pathways lead to photo-oxidation, their investigation on model gas-phase systems may prove useful to elucidate mechanisms of photo-oxidative damages, which are linked to mutations and cancers.

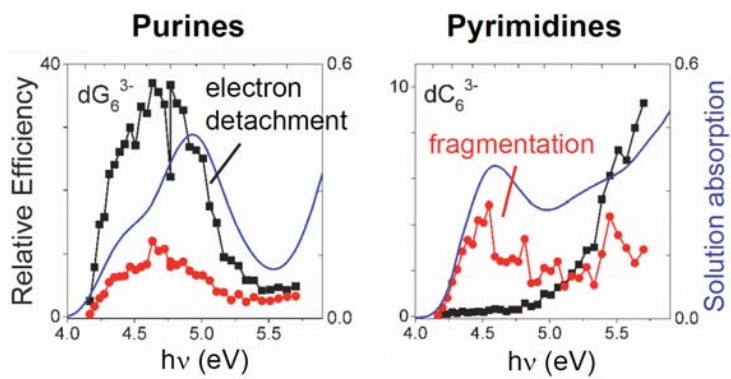

## Introduction

Electrospray ionization is the most convenient way to bring biomolecules, including proteins and nucleic acids, from the solution to the gas phase.[1] The "ionization" in electrospray is a charge separation process, wherein the ions present in solution partition differently in the evaporating charged droplets. As a result, closed-shell singly or multiply charged cations or anions are formed by protonation/deprotonation of basic/acidic groups, or by cation/anion adduction. Proteins and nucleic acids can be analyzed in both polarities, but for nucleic acids, the negative ion mode preserves the solution protonation states: nucleobases remain neutral, and phosphate groups either remain negatively charged like in solution or get neutralized by a proton or cation.

Electrospray ionization allows preserving large biomolecules intact to the mass analyzer. Furthermore, if collisional activation remains moderate during the entire experiment, even non-covalent bonds that were present in solution (especially hydrogen bonds and ionic interactions) can be preserved in the gas phase. Electrospray ionization mass spectrometry in so-called "native" conditions[2] thus opens the possibility to probe the three-dimensional structure of gas phase ions separated in mass, and use the data to infer information about the solution structures of the biomolecules. Native ESI-MS can thus be used to study solution biophysics.

The main gas-phase structural probing techniques are ion mobility spectrometry[3, 4] and ion spectroscopy, and to obtain detailed structures, infrared spectroscopy is often preferred.[5] However, for biomolecules one could envisage using gas-phase electronic spectroscopy in a similar way as UV-vis absorption spectroscopy in solution, with the advantages of pre-sorting mixtures by mass spectrometry and pre-sorting conformations by ion mobility spectrometry. In nucleic acids biophysics, solution UV spectroscopy is routinely used to monitor duplex-to-single strand transitions, folding of guanine-rich strands into G-quadruplexes, or i-motif formation in cytosine-rich strands.[6] Folding is indicated by either a general decrease of absorbance (in case of duplex formation) or by small but significant changes in the shape of the spectra (in case of G-quadruplex formation). The origins of these changes upon structuration are still debated. One surprising property of duplexes is indeed the absence of significant shift in the absorption maximum, despite the delocalization that can occur when chromophores are stacked.[7] Frank-Condon (FC) states in duplex DNA were described as Frenkel excitons,[8] i.e. states involving several base $\pi-\pi^*$ excitations, delocalized over several bases but with no significant charge transfer. These FC states can then (1) efficiently convert back to the electronic ground state via conical intersections,[9] (2) convert into charge transfer (CT) states, from which they can fluoresce,[10] or (3) react to cause photodamage.[11]

Our investigation of electronic ion spectroscopy of nucleic acids was motivated by the possibility of probing the structuration of nucleic acids, and specifically the base arrangements, in the gas phase. In 2006, in collaboration with the Dugourd group we investigated for the first time the interaction of UV photons with DNA multiply charged anions (all containing guanines), and serendipitously found that the main reaction channel was electron photodetachment (ePD), not fragmentation.[12] Furthermore, maximum ePD efficiency was obtained at wavelengths where the bases were known to absorb in solution, suggesting the possibility to use ePD to perform action spectroscopy. In 2007, we reported that the ePD efficiency at 260 nm on $dB_6^{3-}$ ions depends on the nature of the base, and that guanines do particularly favor ePD.[13] We hypothesized that the base-dependent effect was due to differences in electron binding energies. In 2012, we reported the first gas-phase spectra of a 12-base pair duplex and a 4-tetrad G-quadruplex, and compared them to their respective single strands.[14] The structured species showed a significant red-shift compared to the single strands, and we concluded that UV ion spectroscopy was promising to probe nucleic acid structures in the gas phase.

Here we revisit all these results by presenting a new systematic study of 6-mer homo-base single strands ($dG_6$, $dA_6$, $dC_6$ and $dT_6$, see Figure 1), at room temperature and for charge states 2- to 4-. We report their UV ion spectroscopies monitored through fragmentation and electron photodetachment, and show that the action spectrum depends on the action channel (fragmentation or electron detachment), on the charge state, and on the nature of the base. We interpret the results in light of calculated gas-phase structures validated by collision cross sections and infrared multiphoton dissociation (IRMPD) spectroscopy measurements, calculated electronic spectra and vertical detachment energies (VDE), and photoelectron spectroscopy (PES) experiments. These results paint an integrated picture of the inner workings of electronic action spectroscopy on multiply charged anions.

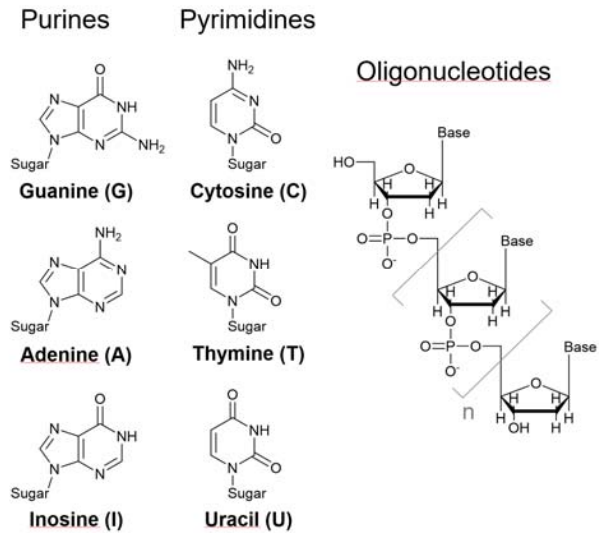

Figure 1. Chemical structures of the purines (guanine, adenine, inosine), pyrimidines (cytosine, thymine, uracil), and structure of the oligonucleotides studied herein (for 6-mers, n=4).

**Materials and methods**

**Samples**

Oligonucleotides were purchased from Eurogentec (Seraing, Belgium) with RP Cartridge-Gold purification, dissolved in nuclease-free water (Ambion, Fisher Scientific, Illkirch, France), and used as is.

**UV absorption spectroscopy in solution**

Absorption spectra have been recorded using a Uvikon XL spectrophotometer. The oligonucleotide concentration was 10 µM and 1 cm optical pathlength suprasil cuvettes were used. The measurements have been performed at 20°C under nitrogen. The bandwidth was 1 nm.

**UV action spectroscopy**

A Spectra Physics PRO-230-30 Nd:YAG laser pumping a GWU Premiscan OPO, running at 30 Hz, was used from 315 nm to 217.5 nm. The laser is guided through a telescope, using an achromatic doublet of focal length 200 mm and a fused silica lens (f=-100 mm). The lenses are separated by 130 mm to obtain a collinear beam. The beam passes through a electro-mechanical laser shutter (SRS 470), which is used to control the number of pulses that will irradiate the ions. Finally, before entry into the mass spectrometer, a thin fused silica blade is used to reflect a portion of the beam to a pyroelectric energy meter (Ophir PE9-C), to monitor precisely the pulse energy during the MS experiment.

The mass spectrometer (AmaZon, Bruker Daltonics, Germany) was modified by Bruker with 1.7 mm diameter holes in the ring electrode and with entrance and exit fused silica windows mounted on the vacuum chamber. After electrospray ionization of 5-µM oligonucleotide aqueous solutions in negative mode, the ions are guided into the ion trap, mass selected with an $m/z$ window of 5 Th, and stored for 70 ms. During this trapping time, the shutter is opened to admit a single laser pulse into the mass spectrometer. The pulse energy is modulated by the alpha-BBO Glan-laser polarizer fixed on motorized precision rotation mount (Thorlabs PRM1Z8). The energy per pulse was adjusted at each wavelength using the polarizing cube. Alignment was checked at each photon energy by verifying we obtained the maximum action yield, and was insensitive to changes in wavelength. In the case of $dC_6^{2-}$, where action yields were not high enough, $dA_6^{4-}$ was co-injected and used for alignment control.

To measure the dependence of the action yields on the pulse energy, the polarizer is rotated in 5-degree steps from 0 to 90 degrees (minimum to maximum transmission), and the pulse energy is measured during the MS acquisition using the reflection from the fused silica blade. For action spectroscopy measurements, the pulse energy is kept at ~160 µJ at each wavelength (although it is reduced at the very extremes of the wavelength range because of absorption of the polarizing cube, see supporting Figure S1). The reflected pulse energy is monitored during each acquisition. Moreover, immediately following an acquisition, both the reflected and transmitted mean pulse energies energies (transmitted pulse energy measured using pyroelectric energy meters (Ophir PE9-C and Ophir PE10-C for the reflected beam and transmitted beam respectively) are recorded for a duration of 30 s (Figure S1). The transmitted pulse energy during an acquisition is thus calculated using Eq. (1), where $E_{trans}^{end}$ and $E_{refl}^{end}$ are the transmitted and reflected pulse energy recorded immediately after an acquisition, and $E_{refl}$ is the reflected pulse energy recorded during an acquisition.

$$E_{trans} = \frac{E_{trans}^{end}}{E_{refl}^{end}} * E_{refl} \qquad (1)$$

After the 70-ms trapping time, the ions are mass-analysed and the mass spectrum is recorded. All acquisitions lasted 2 minutes, which corresponds to a summation of ~400 mass spectral scans. The action yield for each product ion is determined as

$$Y_{product} = \frac{I_{product}}{I_{total}} \tag{2}$$

Where $Y$ is the action yield, $I_{product}$ the integrated area of a product ion peak in the mass spectrum, and $I_{total}$ is the total area (precursor ion plus all the products). We distinguish two classes of products: fragmentation products (i.e., when at least a bond is broken) and electron photodetachment (ePD) products (only electron losses resulting in a change of charge state $z$).

$$Y_{fragm} = \frac{\sum I_{fragm}}{I_{precursor} + \sum I_{fragm} + \sum I_{ePD}} \tag{3}$$

$$Y_{ePD} = \frac{\sum I_{ePD}}{I_{precursor} + \sum I_{fragm} + \sum I_{ePD}} \tag{4}$$

The total action yield is defined as the sum of photofragment yield and electron detachment yield. To reconstruct action spectra, the action yields are normalized by a factor proportional to the number of photons, to obtain the *relative efficiency* (*RE*) of each action:

$$RE_{ePD} = Y_{ePD} / (\lambda * E_{trans}) \tag{5}$$

$$RE_{fragm} = Y_{fragm} / (\lambda * E_{trans}) \tag{6}$$

**IRMPD Action Spectroscopy**

The infrared action spectra of DNA negative ions were studied using an electrospray quadrupole ion trap mass spectrometer (Esquire 3000, Bruker Daltonics, Germany) modified to inject an IR beam in the trap through the ring electrode. All experiments were carried out at the CLIO free electron laser (FEL) center (Orsay, France), which provides an intense and continuously tunable source from 5 to 25 μm with a resolution dλ/λ ≤ 1%. IRMPD spectra are recorded by monitoring the relative fragmentation efficiency (all fragments) of mass-selected parent ions as a function of the excitation wavenumber, in the range 1550-1780 cm$^{-1}$, covering the base NH$_2$ scissoring and C=O stretching modes. The electrospray source parameters were set to minimize ion activation (skimmer -20 V and Cap. Exit at -60 V). Two laser macropulses were used to irradiate the ions (25 Hz, 8 μs width).

**Ion Mobility Spectrometry**

Data were acquired on an Agilent 6560 IMS-Q-TOF (Agilent Technologies, Santa Clara, CA, USA) equipped with the alternate gas kit. The ion mobility cell was operated in helium at 296 K, and the tuning parameters were optimized as described elsewhere.[15] The collision cross sections (CCS) of the centers of the main peaks were obtained using the step-field method ($\Delta V$ = 600, 700, 800, 900 and 1000 V), and the CCS distributions were reconstructed from the arrival time ($t_A$) distributions measured at 600 V, using:

$$CCS = a \cdot \frac{z}{\sqrt{\mu}} \cdot t_A \qquad (7)$$

The parameter $a$ is obtained from the step-field $CCS$ and $t_A$ values of the main peak of each distribution.[16]

**Photoelectron Spectroscopy (PES)**

To obtain sufficient signal, oligonucleotides were prepared at $10^{-4}$ M in water/methanol/ ammonium acetate 100 mM (49:49:2 vol). The experimental setup is described in detail elsewhere.[17] Briefly, ions were accumulated for 1/30 s in a hexapole ion trap, then mass-separated using a reflectron time-of-flight (TOF) mass spectrometer. After mass selection, the ions were impulsively decelerated and then entered the detachment zone of a "magnetic bottle" type of TOF PE spectrometer, where they interacted with the fourth harmonic of a pulsed Nd:YAG laser (4.66 eV) with a pulse duration of 5-6 ns and typical fluence 15 mJ/cm². The typical kinetic energy resolution achieved in this spectrometer is $\Delta E_{kin}/E_{kin}$ < 5% at $E_{kin}$ = 1 eV. The spectra were calibrated against the known photoelectron spectrum of I⁻.

**Molecular Modelling**

*Generation of plausible gas-phase structures.* Generation of candidate structures is challenging because the structures of the single strands are unknown in solution and rearrangements can occur during electrospray. Our goal here was to generate *plausible* structures. Only one phosphate protonation scheme has been used for each strand and charge state. For example, for the 3- charge state, the protons were initially placed on the third and fourth phosphate group (starting from the 5'-end), a choice made based on energy minima in the force field. Several conformers are then generated using temperature replica exchange molecular dynamics (T-REMD, 14 replicas, 2.4 μs), implementing parmbsc1 force field[18] and Amber12 software,[19] and clustered based on their RMSD. Representative conformers of these clusters (provided they had CCS values within the experimental distribution) were then selected for DFT calculations, using Gaussian16 Rev. B.01.[20] Note that DFT enables the protons to be shared or transferred between phosphate groups, and thus the initial protonation scheme is not preserved (see results).

**Simulated IR absorption spectra.** The structures were optimized using the B3LYP functional with the 6-31G(d,p) basis set and an empirical dispersion function Grimme D3, to obtain the theoretical vibrational frequencies. The scaling factor was 0.97. All calculations were performed using Gaussian 16 Rev. A03.

**Simulated UV absorption spectra and calculation of VDE.** For the electronic excitation spectra, the B3LYP structures were reoptimized using M06-2X functional,[21] which is better suited to energetic and electronic calculations on DNA.[22] The basis set was 6-31G(d,p), Grimme dispersion D3, and calculations were performed using Gaussian16 Rev. B.01. TD-DFT calculations were performed by taking into account 200 electronic states. The scaling factor on electronic energies was 0.9. The oscillatory strength of each electronic excitation is converted to molar absorption coefficient (ε) and the simulated spectrum is a combination of the different bands (constructed from Gaussian functions with a peak half-width at half height of 0.33 eV).

**Calculation of collision cross sections.** The values presented in the results section were obtained by trajectory model calculations on the M06-2X optimized structures, in helium at 300 K, performed using the Mobcal software[23] using its original parameterization (P, N and O sharing the same parameters as carbon).

## **Results**

### *UV action channels*

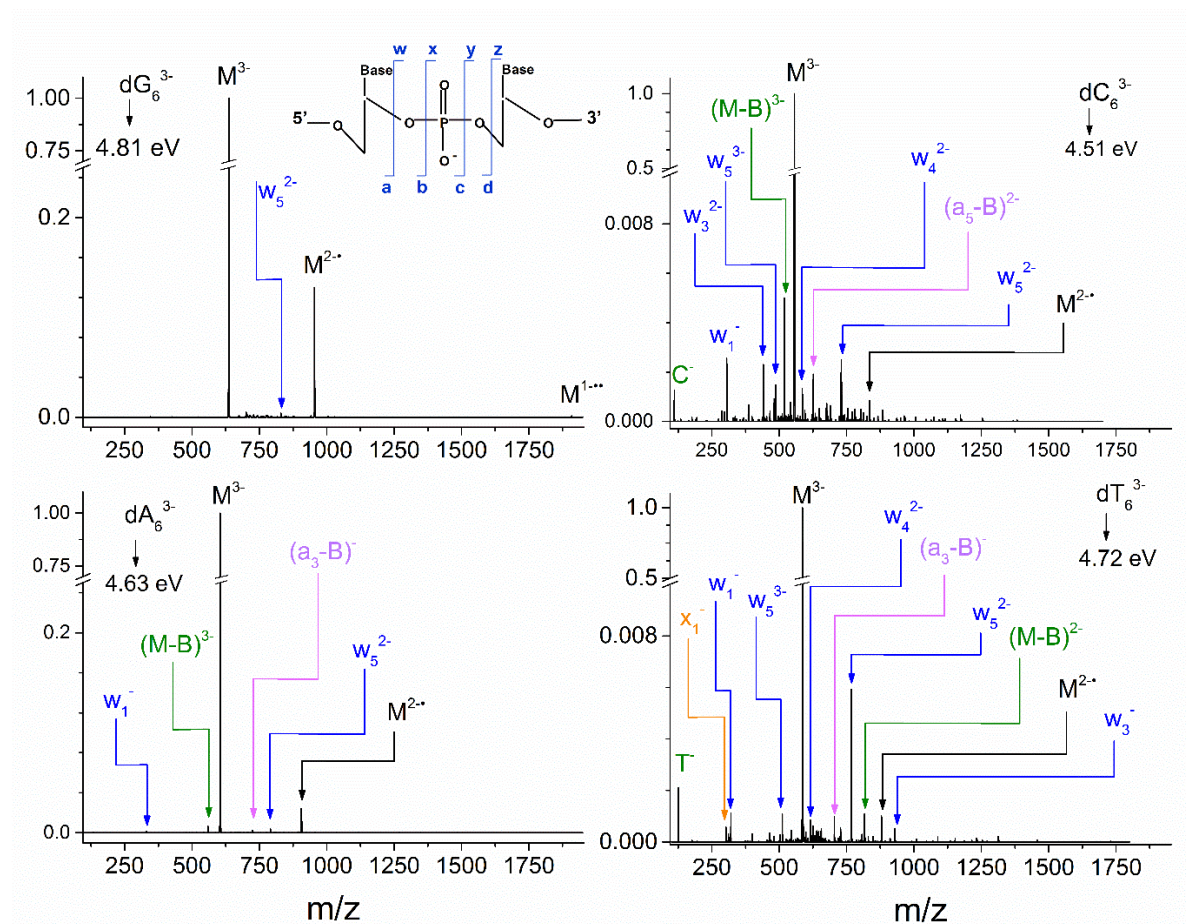

Figure 2. Mass spectra following irradiation with a single UV laser pulse (160 µJ transmitted through the trap) at 257.5 nm (4.81 eV) for $dG_6^{2-}$ (top left), 275 nm (4.51 eV) for $dC_6^{3-}$ (top right), 267.5 nm (4.63 eV) for $dA_6^{3-}$ (bottom left)and 262.5 nm (4.72 eV) for $dT_6^{3-}$ (bottom right). Major fragmentation channels are labelled according to the nomenclature in the inset, and ePD products are shown in black. Note that the vertical scale for purines (A and G) differs from that of the pyrimidines (C and T).

Figure 2 shows the mass spectra of $dA_6^{3-}$, $dC_6^{3-}$, $dG_6^{3-}$ and $dT_6^{3-}$ irradiated with a single laser pulse of 160 µJ at the solution absorption maximum of each base (data for the 2- and 4- charge states are in supporting Figures S2 and S3). For the purines, electron detachment is the dominant photoreaction channel for the 3- charge state, and $dG_6^{3-}$ detaches electrons more efficiently than $dA_6^{3-}$, as reported previously at 260 nm.[14] Electron detachment is however a minor channel in both $dC_6^{3-}$ and $dT_6^{3-}$. These strands show extensive fragmentation instead, with *w*, *a-B* and base loss being the most important fragmentation channels (see inset for nomenclature[24]). For both purines and pyrimidines, the same photofragments are observed with collision-induced dissociation (see supporting Figures S4—S15), with the exception of some neutral losses from the radical anions produced by electron detachment.

### *How action yields depends on laser pulse energy*

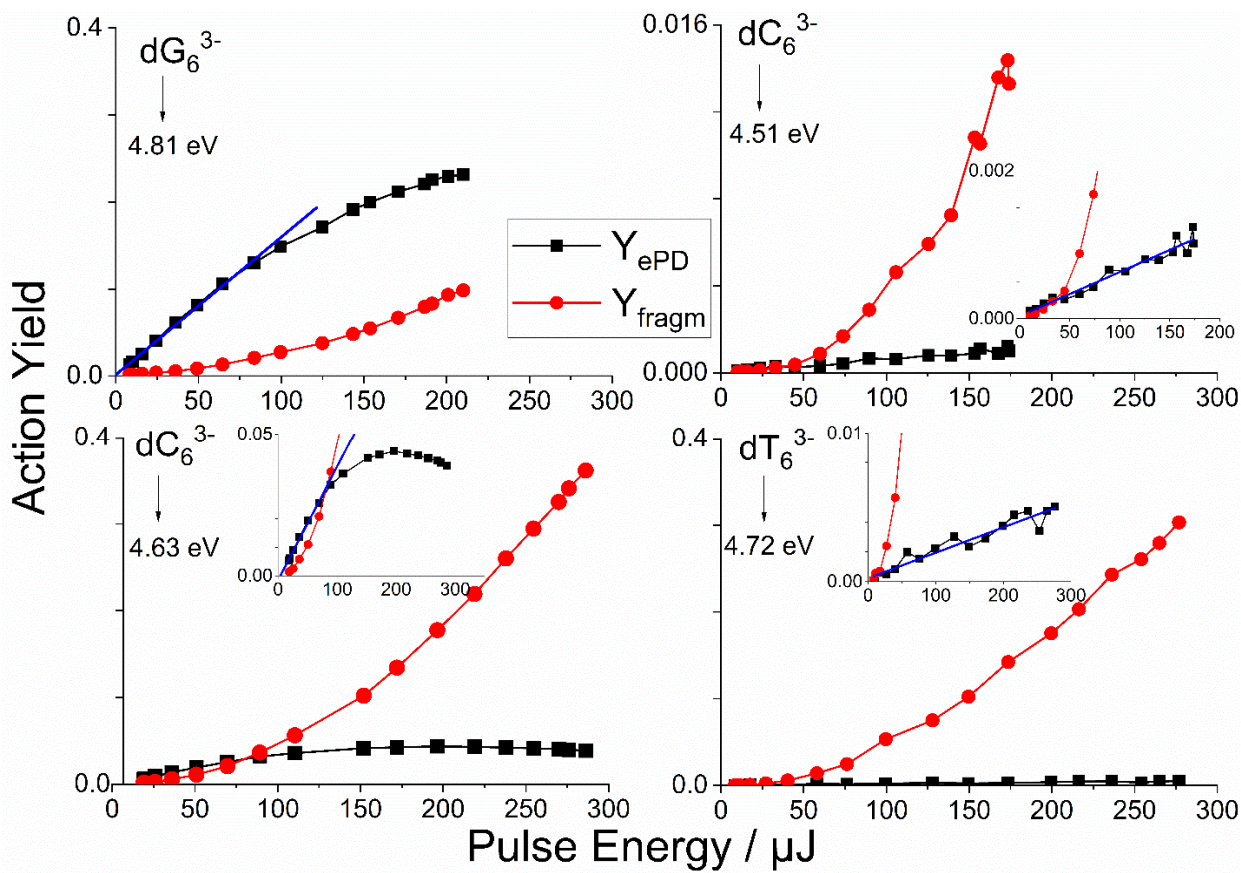

Figure 3. Photoreaction yields versus transmitted pulse energy $E_{trans}$ for ePD (black) and PF (red) for the 3- charge states of $dG_6$ (top left, 257.5 nm, 4.81 eV), $dA_6$ (bottom left, 267.5 nm, 4.63 eV), $dC_6$ (top right, 275 nm, 4.51 eV) and $dT_6$ (bottom right, 262.5 nm, 4.72 eV). Insets show zooms of the same data. The blue lines are a linear fit of the ePD data.

Figure 3 shows how electron photodetachment and the fragmentation yields depend on the pulse energy for $dA_6^{3-}$, $dC_6^{3-}$, $dG_6^{3-}$ and $dT_6^{3-}$ at their respective predicted absorption maxima. The results for the 2- and 4- charge states are shown in supporting Figures S16—S17. The ePD yield increase linearly for all species (blue lines) and the intercept is (0,0), although for the purines the yield levels off at high pulse energies due to secondary absorption and photoreaction of the charge-reduced anion. The fragmentation yield is characterized by a nonlinear dependence on the pulse energy. The same holds for the individual fragments (supporting Figures S18—S20). The linear dependence of the electron detachment yield shows that ePD is a one-photon process (as observed previously on other polyanions[13, 25-27]), whereas the nonlinear increase of the photofragmentation yield indicates a multiphoton process.

The oligopurines show higher ePD yields than the oligopyrimidines. For $dG_6$, ePD predominates at all pulse energies. For $dA_6$, ePD predominates below ~80 µJ, and photofragmentation dominates at higher pulse energies. For pyrimidines, fragmentation predominates at all pulse energies. Further, the total action of $dC_6$ is an order of magnitude lower than for any other oligonucleotide. Similar trends are observed in the other charge states, see supporting Figures S16—S17, with the exception that, for $dG_6^{2-}$, fragmentation is more efficient than electron detachment above 55 µJ.

**UV action spectroscopy**

All UV action spectra are shown in Figure 4. The trend in base-dependent ePD yields is generally conserved at different charge states and wavelengths. At photon energies corresponding to the first solution absorption band, the ePD efficiency ranks G >> A >> T ~ C. The latter two (the oligopyrimidine strands) actually show no proper "band" in the ePD action spectra below 5 eV (by "band" in the action spectra, we mean that the action efficiency increases, then decreases significantly when the photon energy is increased). We also notice that, among the 2- charge states, only $dG_6^{2-}$ shows a band in the ePD action spectrum.

In contrast, the photofragmentation action spectra all show bands in the wavelength region corresponding to the first solution absorption maximum (4.6—5.0 eV, i.e. 248—270 nm). Although the shapes are not identical, the bands overlap well with the solution absorption ones. Because fragmentation is a multiphoton process, while the yield is linearly normalized by the photon flux, the same shape is not expected, but we can compare the position of the action maxima. The action spectra's maxima (both for fragmentation and for ePD when observed) are generally shifted to lower photon energies (red-shifted) compared to the solution ones, except for $dG_6^{2-}$ (ePD action spectrum) and $dT_6^{2-}$ (fragmentation action spectrum) which follow the solution absorption spectra very well, and $dC_6^{2-}$ which is slightly shifted to higher photon energy. For $dA_6$, $dT_6$ and $dC_6$, the shift is relatively modest (~0.1 eV), but for $dG_6^{3-}$ and $dG_6^{4-}$ the red-shift of the maximum is more pronounced (~0.25 eV).

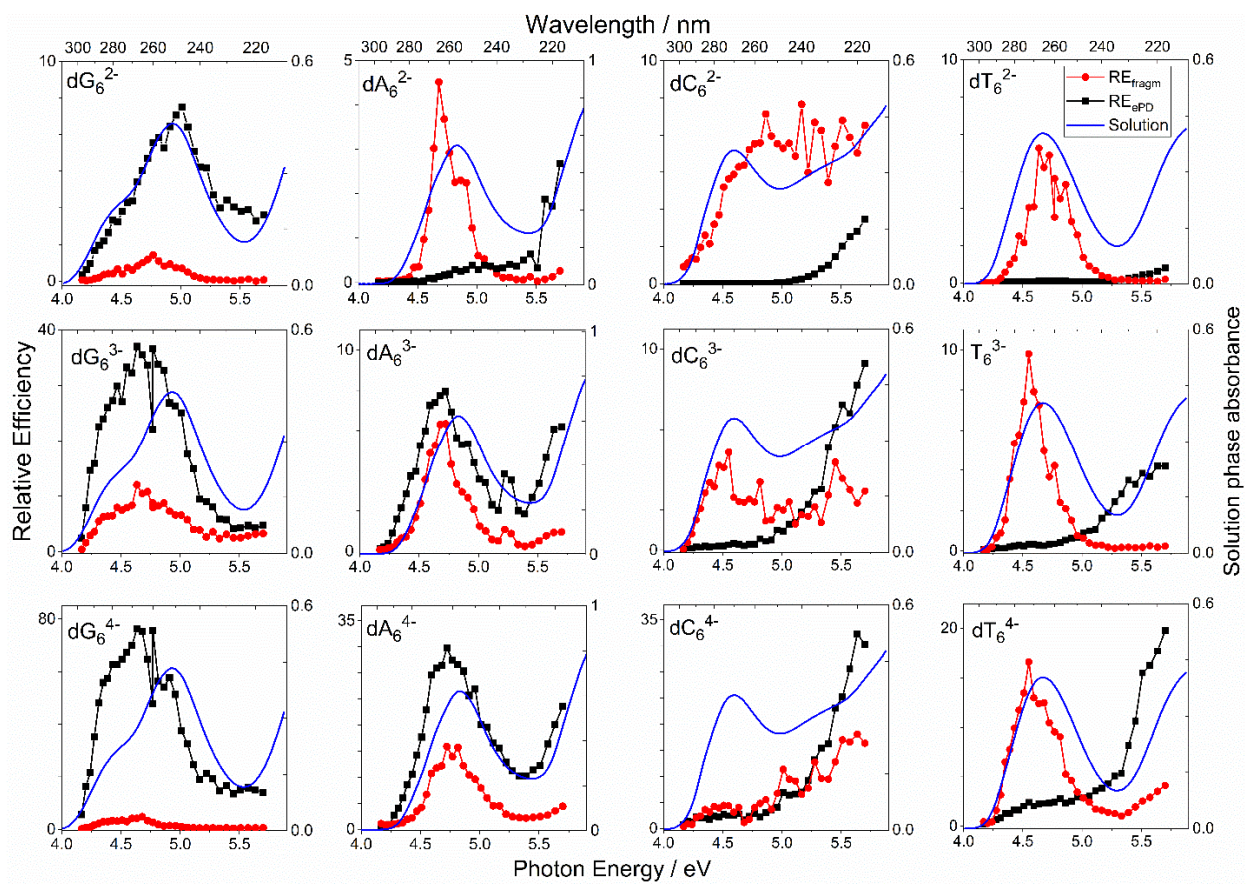

Figure 4. UV action spectra for oligonucleotides (from left to right) $dG_6$, $dA_6$, $dC_6$ and $dT_6$, and for charge states (top to bottom) 2-, 3- and 4-. The spectra reconstructed based on electron photodetachment relative efficiency are in black, and the spectra reconstructed based on fragmentation are in red. The solution absorption spectra are superimposed.

In solution, the absorption increases again above 5.5 eV (below 225 nm). The same trend is also found in the ePD action spectra (except for $dG_6$), for all charge states. In the photofragmentation action spectra the trend is less systematic, and this may be due to the combined effects of the decaying photon flux at these wavelengths and of the multiple-photon character of photofragmentation.

**Gas-phase structures and predicted gas-phase absorption spectra**

The gas-phase structures were probed by ion mobility spectrometry and infrared multiphoton dissociation (IRMPD) action spectroscopy, as illustrated in Figure 5 for $dG_6^{3-}$. All collision cross section distributions are shown in Figure 6. In most cases, the gas-phase structures consisted of several non-interconverting structures. The proton locations for the different conformers are summarized in Figure S21, and we see that the generated conformers are in general also protonation isomers (despite the proton attachment sites before optimization were the same). The 4- charge states show extended conformations, indicating that Coulomb repulsion has a major effect on the structure. However, the conformations of $dG_6$, $dA_6$ and $dC_6$ are significantly more compact in their 3- and 2- charge states, indicating that other intramolecular interactions overcome part of the Coulomb repulsion. We then examined the base-related region (1550—1770 $cm^{-1}$) by IRMPD spectroscopy. The band positions reflect the average degree of hydrogen binding in which the bases are involved. All spectra are shown in supporting Figure S22.

To select plausible gas-phase structures on which to calculate electronic absorption spectra, we examined how well both the collision cross section values and the calculated harmonic IR absorption spectra match with the experimental data, as illustrated for $dG_6^{3-}$ in Figure 5 (the IRMPD spectra for the other strands and charge states are shown in supporting information Figures S23—S25, and the matching between theoretical and experimental CCS for all strands is shown in Figure S26).

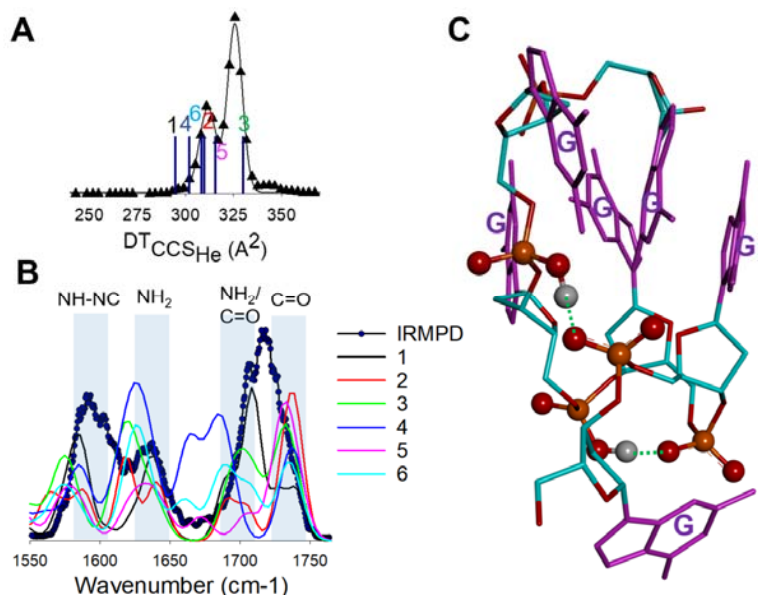

Figure 5: A) Ion mobility spectrometry of the single strand dG$_6^{3-}$ ($^{DT}$CCS$_{He}$). The vertical bars are the theoretical collisional cross section obtained on the different conformers of dG$_6^{3-}$ using the trajectory model (TM, mobcal). B) Experimental IRMPD spectrum of the single strand dG$_6^{3-}$ and the calculated IR spectra of different conformers of dG$_6^{3-}$. The calculations were performed using Gaussian 16 rev. B01 (DFT, B3LYP, 6-31G(d,p)+GD3) and frequencies were scaled by a factor of 0.97. C) Structure of conformer 5 of dG$_6^{3-}$. Hydrogen bonds between two pairs of phosphate groups are represented in ball and stick.

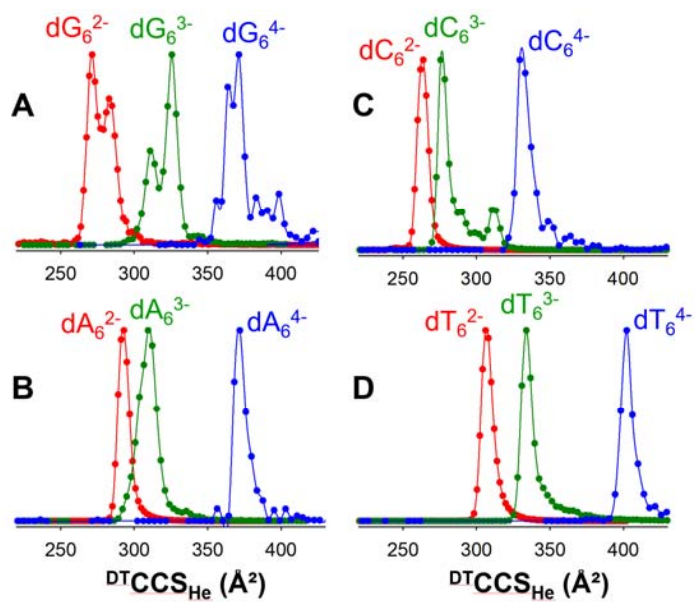

Figure 6. Collision cross section distribution obtained in a helium drift tube at 296 K for the electrosprayed single strands at different charge states. A: dG$_6$, B: dA$_6$, C: dC$_6$, D: dT$_6$.

Conformers labeled 2, 3, 5 and 6 fit within the experimental CCS distribution, and all conformers except 4 have IR spectra matching with the IRMPD experiment. The structure of conformer 5 is shown in panel 5C: the strand is folded thanks to the formation of hydrogen bonds between phosphate groups (the protonation scheme changed from 5'-00110-3' to 5'-½½½½0-3'). Groups of bases are stacked (a group of three, a group of two). Moreover, hydrogen bonds are formed between the $NH_2$ group or NH group of the different guanine and phosphate groups. For the different conformers, C=O stretching modes are found between 1720 cm$^{-1}$ and 1750 cm$^{-1}$. The region between 1620 cm$^{-1}$ and 1650 cm$^{-1}$ corresponds to vibrational modes involving $NH_2$ scissoring mode. The 1670—1720 cm$^{-1}$ corresponds to a population of highly concerted vibrational modes (C=O stretching and $NH_2$ scissoring) from bases involved in hydrogen bonds (between base-base or base-phosphate): the C=O stretching is red-shifted and the $NH_2$ scissoring is blue-shifted (see animated Figure S27 in supporting information).

For each oligonucleotide in the 3- charge state, we calculated the UV absorption spectra of several conformers in the gas phase by TD-DFT, and the results (Figure S28) show that the gas-phase absorption spectra do not depend much on the conformer choice. Thus, on Figure 7, we show the calculated action spectrum of one conformer (matching the ion mobility and IRMPD data) for each charge state of the various oligonucleotides. When an offset of 0.9 is applied to the photon energies (which is typical at this level of theory[28, 29]), the calculated spectral shapes match very well with the solution ones, including with respect to peak intensities (molar extinction coefficient). It is possible that the scaling factor is wrong, and that all gas-phase spectra are actually shifted compared to the solution spectra. However, we note that the charge state does not significantly influence the gas-phase absorption properties, so we do not anticipate that the gas-phase spectra should differ greatly from the solution absorption spectra. We thus think that comparing the gas-phase action spectra to the solution absorption spectra, as we did in Figure 4, is a good basis for discussing whether the gas-phase action spectra reflect the absorption spectra.

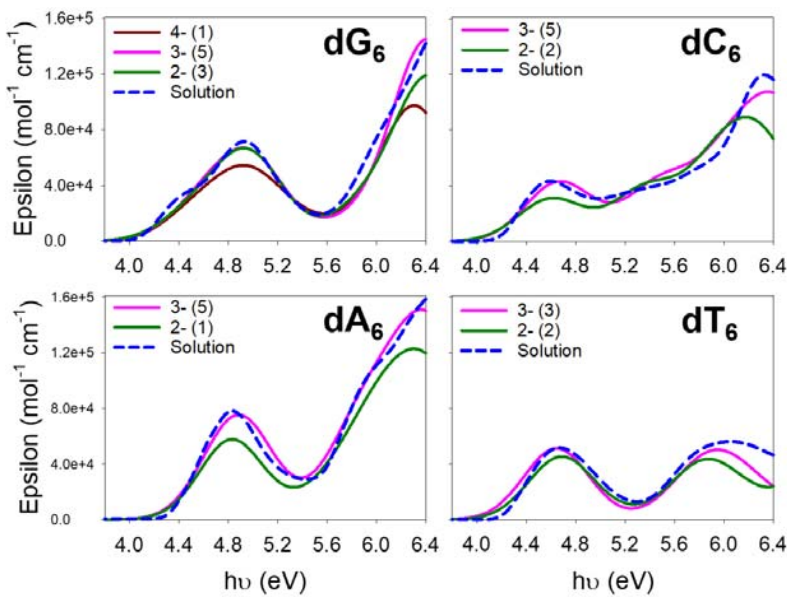

Figure 7. Calculated absorption spectra (TD-DFT/MO6-2X/6-31G(d,p),GD3), hv values scaled by 0.9) in the gas phase (solid lines) compared with solution absorption spectra (dashed blue lines) of the single strand $dA_6$, $dC_6$, $dT_6$ and $dG_6$. For each charge state (pink for 3-, green for 2-, brown for 4-) one representative conformer has been selected (conformer number indicated in the legend, matching with the annotations in the IRMPD spectra, Figures S23—25, and the CCS distributions, Figure 6).

**Electron binding energies**

We saw that the ePD action spectra (and overall ePD efficiencies) depend on the base nature and on the charge state. For 1-photon ePD to be possible, the photon energy must be higher than the adiabatic detachment energy (ADE), i.e. the energy difference between the precursor ion and the corresponding radical, see Figure 8. However, in multiply charged anions, an additional barrier, called the repulsive Coulomb barrier (RCB), is present, because an electron cannot approach the anion radical with infinitely low kinetic energy and bind to it; it would be repelled by Coulomb repulsion. Electrons can still tunnel through the RCB, with low efficiency, but high ePD efficiencies can only happen if hv > ADE + RCB. We had previously attributed the base-dependent detachment efficiencies to putative differences in the electron binding energies.[13] Here we tested this hypothesis using photoelectron spectroscopy (PES) data and vertical detachment energy (VDE) calculations.

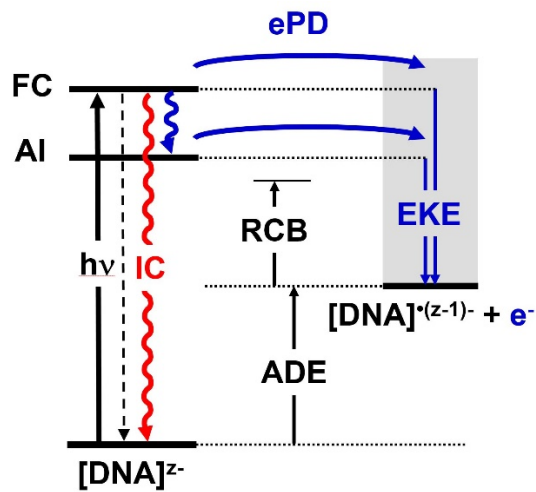

Figure 8. Energy diagram for UV action spectroscopy and photoelectron spectroscopy on DNA multiply charged anions. A photon (hν) resonantly excites [DNA]$^{z-}$ to a Frank-Condon states (FC). The system can relax back to the electronic ground state either by photon emission (dashed black arrow), or by internal conversion (IC) of electronic energy to vibrational energy. For electron detachment to be possible, the photon energy must be higher than the adiabatic detachment energy (ADE) plus the repulsive Coulomb barrier (RCB). The blue arrows show the electron kinetic energies (EKE) expected with a direct detachment from the FC states, and with direct detachment from a putative auto-ionizing (AI) state of the precursor ion.

Electron binding energies were measured using photoelectron spectroscopy.[30] Figure 9 shows the effect of the base nature on the 3- charge state of the 6-mers. In line with the ePD action spectra presented in Figure 4, the electron yields were much higher for dG$_6$$^{3-}$, followed by dA$_6$$^{3-}$. The yields of dC$_6$$^{2-}$ and dT$_6$$^{3-}$ were comparatively very low, but non-zero. The thresholds at which ePD increased significantly corresponded to adiabatic detachment energies (ADE, ± 0.2 eV, in red in Figure 9, in black in Figure 10) of ~1.85 eV for dG$_6$$^{3-}$, ~2.0 eV for dA$_6$$^{3-}$, ~2.35 eV for dC$_6$$^{3-}$, and ~2.5 eV for dT$_6$$^{3-}$ (although for the latter two, a shallow contribution is visible down to 2.0 eV).

The values of the RCBs can be estimated from the electron kinetic energies (EKE = 4.66 – ADE on Figure 9) at which electron detachment is significantly suppressed. Given that the RCB can depend on the electronic excited state and the ion conformation, we estimate here its maximum value. By estimating the EKE until the last intense feature of the PES spectra (green arrows), one obtains a maximum RCB of 1.55 eV for dG$_6$$^{3-}$ and dC$_6$$^{3-}$, 0.95 eV for dA$_6$$^{3-}$, and 1.15 eV for dT$_6$$^{3-}$

. The threshold for efficient electron photodetachment (ADE + RCB) is therefore maximum ~3.4 eV for $dG_6^{3-}$, ~2.95 eV for $dA_6^{3-}$, ~3.9 eV for $dC_6^{3-}$, and ~3.65 eV for $dT_6^{3-}$.

Two bands in the PES are common to $dG_6^{3-}$, $dA_6^{3-}$ and $dC_6^{3-}$ (in blue in Figure 9), at EBEs = 2.61 and 2.91 eV, i.e. corresponding to EKEs of 2.05 eV and 1.75 eV. They are absent in $dT_6^{3-}$, but interestingly, as soon as a single guanine is inserted in a poly(thymine) strand (see supporting Figure S29),[30] these hot bands reappear, and the ADE drops (from ~2.5 to ~1.85 eV), as though the guanine alone was responsible for most emitted electrons.

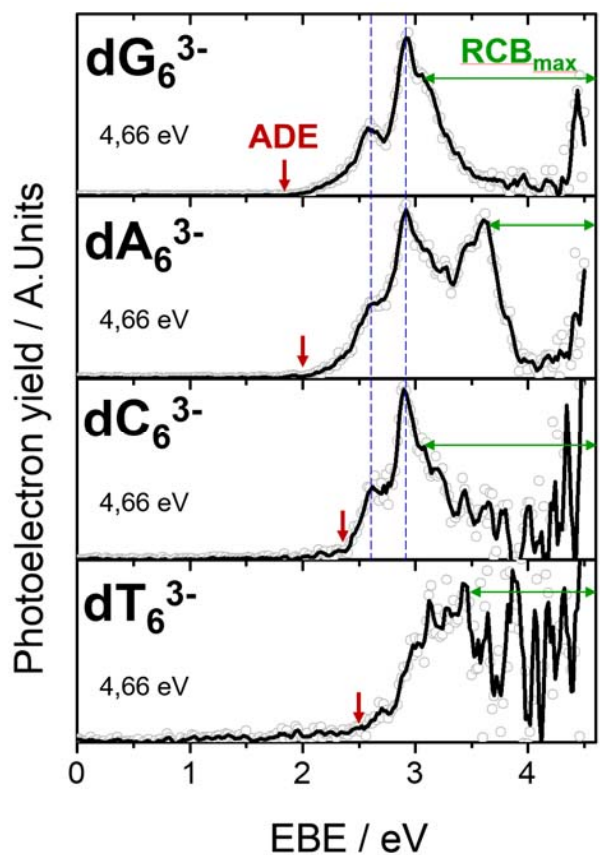

Figure 9. Photoelectron spectra of $dG_6^{3-}$, $dA_6^{3-}$, $dC_6^{3-}$ and $dT_6^{3-}$, with an excitation wavelength of 266 nm (hν = 4.66 eV).

The influence of the charge state on the electron binding energies was investigated for the strand dGGGTTT (see Figure S30).[30] The ADE values are -0.2 eV for the 5-, 0.6 eV for the 4-, 1.6 eV for the 3-, and 3.2 eV for the 2-. Negative ADEs for fully deprotonated oligonucleotides were previously reported for $dT_5^{4-}$,[31] $dA_5^{4-}$,[31, 32] $dG_5^{4-}$,[32] $dC_6^{5-}$.[30] The exact binding energies depend on the conformer or protonation isomer.[32]

We also calculated the vertical detachment energies (VDE) for the different strands, different charge states, and for several conformers. The results are shown in Figure 10. The data for the 3- ions are distributed right above the ADE values determined by PES, and thus agree well with the experiments. The VDE values for the 2- ions lie on average 1.79 eV (standard deviation: 0.25 eV) higher than those of the 3- ions, which agrees with the ADE increase from 3- to 2- in dGGGTTT.

In summary, the results show that for all 3- and 4- ions investigated here, the photon energies used in action spectroscopy (minimum 4.1 eV) are all higher than ADE + RCB. However, for the 2- ions, it is possible that the lowest photon energies may lie under the repulsive Coulomb barrier.

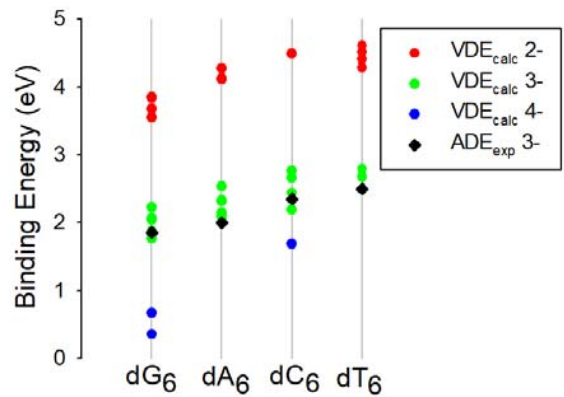

Figure 10. Calculated vertical detachment energies (TD-DFT/MO6-2X/6-31G(d,p),GD3) for the hexadeoxynucleotides 2- (red), 3- (green), and 4- (blue). The adiabatic detachment energies extrapolated from the photoelectron spectroscopy experiments of Figure 9 on the $dB_6^{3-}$ ions are shown in black.

**Discussion**

1) **Fragmentation channels and their use for UV action spectroscopy**

Since the photofragmentation channels resemble those observed upon low-energy collision (pseudo-thermal) activation, we will assume that photofragmentation is the consequence of internal conversion (IC) of the electronic energy into vibrational energy, followed by complete intramolecular vibrational energy redistribution (IVR). Photofragmentation of 6-mer oligodeoxynucleotides requires multiple photons (as opposed to a single photon for mononucleotides[33]) and can be coined UVMPD (ultraviolet multiple-photon dissociation), by analogy with IRMPD. The multiple-photon character is due to the number of vibrational degrees of freedom onto which the internal energy redistributes. The typical threshold for base loss (which then triggers fragmentation[24]) is around 1 eV,[34] and our 6-mers comprise > 540 vibrational degrees of freedom. A single photon having 4 to 6 eV of energy is not sufficient to fragment a 6-mer oligonucleotide.

For photofragmentation to occur, the energy must first have been deposited in the molecule by resonant excitation to reach electronic excited states. This is why the action spectra reconstructed from the photofragmentation channels generally follow the expected absorption spectra. However, the multiple-photon character of the process makes it difficult to infer the absorption efficiency directly from the action efficiency. One does indeed not know *a priori* how many photons are required to reach a certain level of fragmentation for a given channel, so one does not know how to normalize the action yield to reflect the absorption yield. Here, to reconstruct the action spectra, we normalized the yield linearly by the number of photons, but this is arbitrary. A consequence is that, for large molecules fragmenting via UVMPD, the shape of the action spectra reconstructed based on fragmentation will depend not only on the molecular system, but also on the photon density.

Another case where the UVMPD action spectra may not strictly follow the absorption spectra is when the excited states react through other channels than IC. The probability of a given excited state to proceed to IC also depends the competition with other relaxation channels, for example photon emission (fluorescence) or electron emission (electron detachment). Fluorescence typically has very low (~$10^{-4}$) quantum yields in oligodeoxynucleotides,[35] while internal conversion is very efficient and fast (low picosecond range).[9] However, electron detachment is observed in our experiments, meaning that it effectively competes with IC. In turn, this means that ePD is also very fast. Let's now examine how ePD may be related to absorption.

## 2) Electron photodetachment (ePD) and its use for UV action spectroscopy

### 2.1. Base-dependent effects are not fully explained by electron binding energies

Electron photodetachment of oligonucleotide polyanions is a single-photon process. Our action spectra were recorded at $h\nu \geq 4.1$ eV, and the first resonance absorption bands are expected at ~4.5 eV. For 1-photon ePD to be *possible*, the photon energy ($h\nu$) must be higher than the adiabatic detachment energy (ADE), and for 1-photon ePD to be *efficient*, the photon energy must be higher than the ADE augmented by the repulsive Coulomb barrier ($h\nu >$ ADE + RCB).

For the 3- charge state, the entire range of $h\nu$ is entirely above ADE + RCB, whatever the nature of the base. *A fortiori*, the same will hold for all 4- ions. Yet efficient ePD in the base absorption region is only observed in the case of $dG_6$ and $dA_6$. The ePD efficiency for the 3- and 4- charge states, which is grossly G >> A >> T ≈ C (although wavelength-dependent), does not follow the electron binding energies, and thus electron binding energies alone do not explain the base-dependent photodetachment efficiencies.

### 2.2. Charge state effects on the ePD action spectra

For the 2- charge state, given the ADE measured for the strand $dGGGTTT^{2-}$ and the VDEs calculated for our $dB_6^{2-}$ strands, it is plausible that, in the high-wavelength part of the spectrum (low photon energies), $h\nu <$ ADE + RCB. The ePD action spectrum of $dG_6^{2-}$ (Figure 4) is skewed, with an apparent linear increase of the ePD efficiency starting at $h\nu_{onset,ePD} \approx 4.1$ eV. For $dA_6^{2-}$, $h\nu_{onset,ePD} \approx 4.55$ eV. The average calculated VDE for $dG_6^{2-}$ is $\overline{VDE_{calc}} = 3.72 \pm 0.12$ eV (standard deviation from all conformers) and for $dA_6^{2-}$, $\overline{VDE_{calc}} = 4.17 \pm 0.07$ eV. The energy differences between the two strands are the same, thus one possibility is that the $h\nu_{onset,ePD}$ indicates the position of the lowest possible repulsive Coulomb barrier.

Note however that if ADE < hν < ADE + RCB, electron detachment is still possible, but the electron then has to tunnel through the RCB. This process would be less efficient than if hν > ADE + RCB, and the efficiency would progressively increase as hν increases, because the barrier gets narrower. The other possibility is thus that, for $dA_6^{2-}$ and $dG_6^{2-}$, the linear increase starting at $h\nu_{onset,ePD}$ corresponds to hν < ADE + RCB. This means that, because of the repulsive Coulomb barrier, the bands in the ePD action spectra of $dA_6^{3/4-}$ and $dG_6^{3/4-}$ are either suppressed in $dA_6^{2-}$, or skewed in $dG_6^{2-}$.

An important consequence is that the ePD action spectra are charge-state dependent, even when the absorption spectra are not, because if too low charge states are used, the photon energy may lie too close to the ADE, and ePD may be inhibited in a wavelength-dependent manner. Consequently, before comparing the ePD action spectra on different molecular systems, especially if they have different sizes, and thus charge densities and RCBs, the charge-state dependence of ePD should be investigated. Comparing one charge state of each, as we did in a previous report,[14] does not provide a complete enough picture to interpret ion spectroscopy in terms of structure. To highlight this, we reproduced the study comparing the tetrameric G-quadruplex [dTGGGGT]$_4$ and the single dTGGGGT, and studied three charge states for each structure (Figure 11). The red-shift of the G-quadruplex compared to the single strand (Fig. 11A) is much less obvious, because a charge state increasing also produces the same kind of red-shift (Fig. 11B—C).

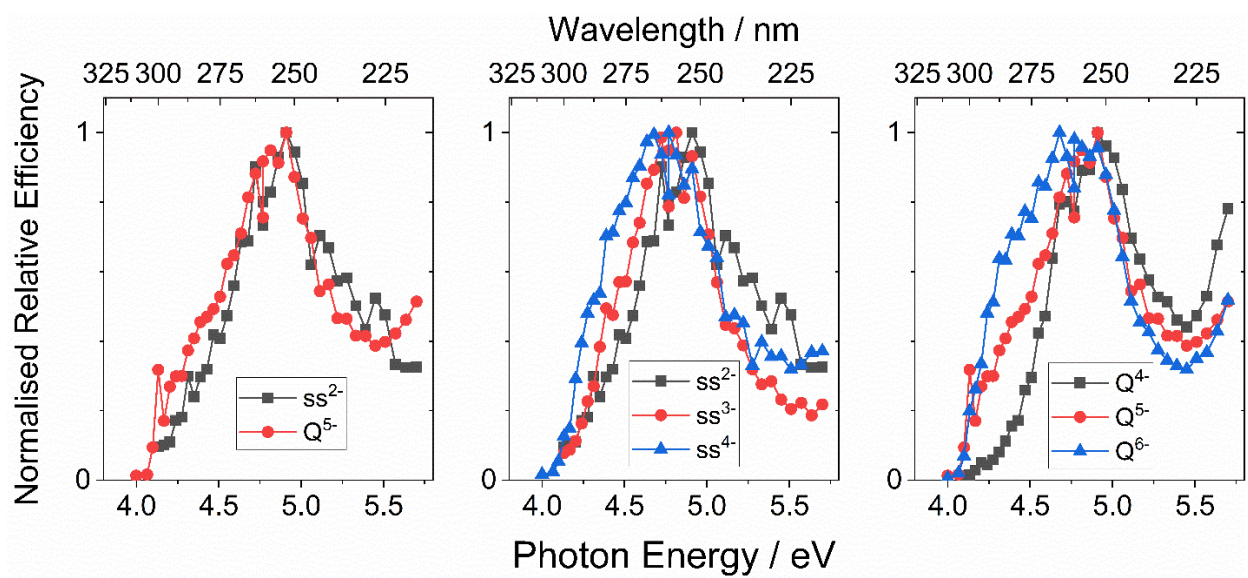

Figure 11. ePD action spectra of the G-quadruplex [(dTGGGGT)$_4$(NH$_4^+$)$_3$] (Q) and of the corresponding single strand dTGGGGT (ss), at different charge states. For both systems, the shape of the action spectra changes with the charge state.

## 2.3. Purines vs. pyrimidines

For the oligopyrimidine (C and T) strands, the ePD efficiency increases progressively with hν, but no resonance band is observed at the first absorption maximum around 4.6 eV, whatever the charge state. This is not due to the absence of absorption, because in contrast to ePD, the maximum fragmentation yield occurs around the expected absorption maximum. To confirm this trend, we performed action spectroscopy experiments on one more homopurine strand (uracil; dU$_6^{4-}$) and one more homopyrimidine strand (inosine; dI$_6^{4-}$). The results (Figure 12) confirm the trend: a resonance band is observed in the ePD action spectra of dI$_6^{4-}$, and not in dU$_6^{4-}$, for which photofragmentation occurs instead. The stark contrast between the behavior of the purines and the pyrimidines remains unexplained by differences in absorption or by differences in electron binding energies.

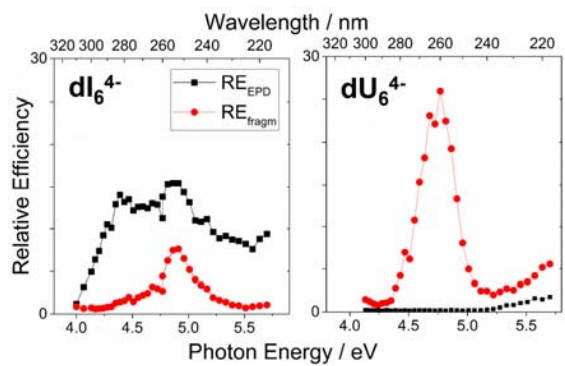

Figure 12. ePD action spectra (black) and photofragmentation action spectra (red) for the polypurine strand $dI_6^{4-}$ (left) and the polypyrimidine strand $dU_6^{4-}$ (right).

Thus, the condition that $h\nu > ADE + RCB$ is not sufficient to ensure that the ePD action spectrum reflects the absorption efficiency. The electronic excited states must also have special properties, which purines studied so far have and pyrimidines studied so far don't have. It is also possible that not all excited states of purines have the same propensity to ePD. For example, we notice that the ePD action spectra of $dG_6^{3-}$ and $dG_6^{4-}$ are significantly shifted to lower-energy compared to the expected absorption spectra, with a notable band at ~4.4 eV (Figure 4). It is possible that, at this energy, lie electronic states that are particularly favorable to ePD.

We don't know yet what these special excited states properties could be. Here we studied relatively unstructured single strands. The bases are not neatly stacked but neither are they independent of each other. Our calculations show that the first electronic transitions of high oscillatory strength, at energies corresponding to the first ePD band, involve molecular orbitals that are delocalized on several bases (Figure 13 for $dG_6^{3-}$), akin to Frenkel excitons. This was however obtained for purines and pyrimidines alike (see supporting information Figures S31—33 for the other 6-mers[3-]), so the "delocalization" of the excited states reached first upon photon absorption (the bright states) is not explaining the difference.

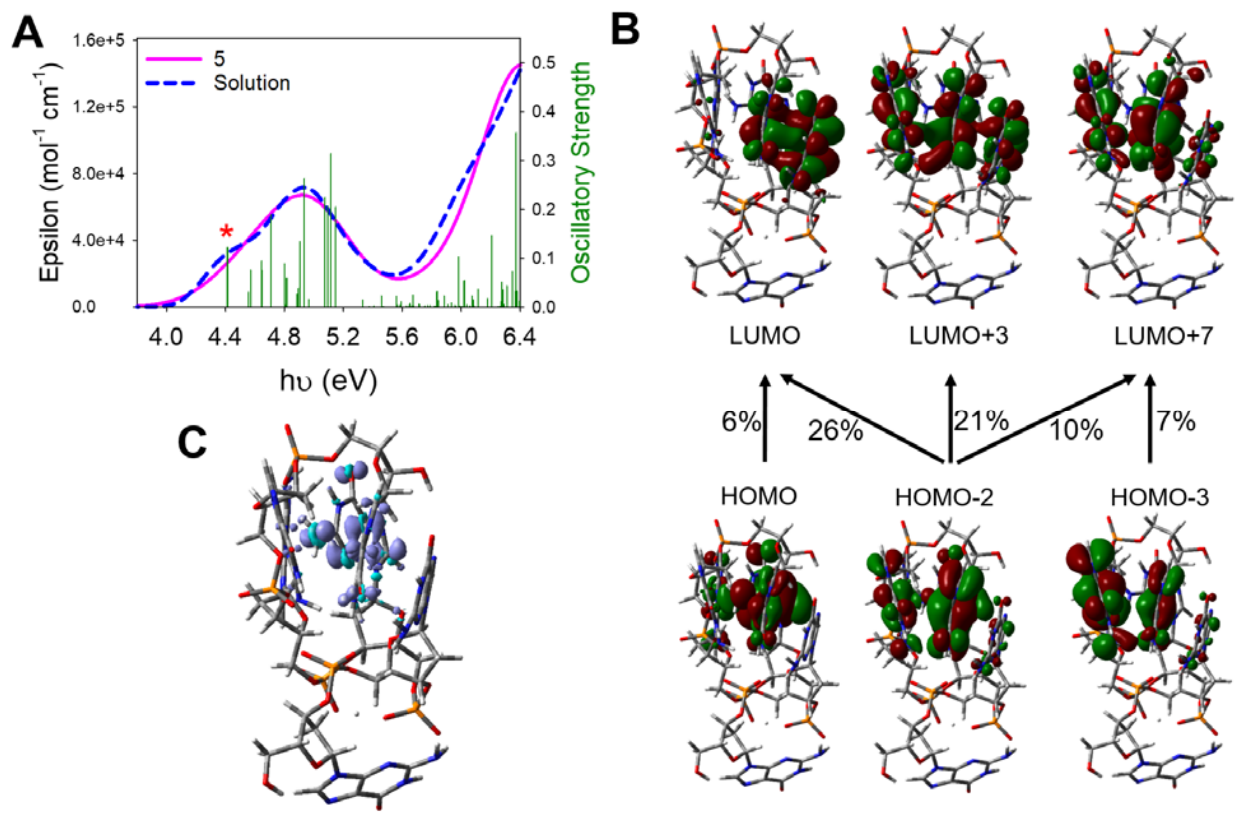

Figure 13. A) Experimental solution absorption spectrum and calculated gas-phase absorption spectrum (conformer 5) of dG$_6^{3-}$. B) Molecular orbitals involved in the first electronic transition of significant oscillator strength (shown by a star in panel A). C) Calculated difference of electronic densities between dG$_6^{3-}$ and the product of vertical electron detachment, dG$_6^{2-•}$. The difference (electron loss) is located on the bases.

Photodetachment may also proceed not directly from the initially populated states, but from other (dark) excited states to which the initially populated states relax first. Antoine and Dugourd proposed a similar mechanism for the ePD of peptides and proteins, and coined these states "auto-ionizing states".[36] Electronic excited states with electron density transferred in σ* orbitals (πσ* states), where electrons are thus moved further away from the nucleobase in orbitals with substantial Rydberg character,[37] are potential candidates to serve as auto-ionizing states. Such states have been described for guanine,[38] adenine,[39, 40] and cytosine.[28] While we note that oligonucleotides may not behave like isolated bases, it remains possible that such dark excited states of purines are favorable to ePD (are auto-ionizing), whereas those of pyrimidines either do not exist, or are not auto-ionizing. Another possibility is that charge transfer (CT) states (excimers or exciplexes[41, 42]), which have a parentage with the FC states,[35] may serve as auto-ionizing states. As illustrated in Figure 8, efficient auto-ionizing states should result in "hot bands" in the PES spectra, at particular EKE values. The energy of the AI states would thus be equal to the ADE + the EKE of the hot bands. In Figure 9, hot bands are observed at EKE = 2.05 eV and 1.85 eV (shown in blue). With an ADE of ~1.85 eV, this gives AI states ~3.7 eV and ~3.9 eV above the electronic ground state, i.e. in the region 320—340 nm, where DNA fluorescence is typically observed in solution.[35] Thus, charge transfer states such as those involved in flurescence could also be involved in ePD, with high efficiency for purines.

Finally, to have more insight from where an electron can be removed, we have calculated the difference of electronic density between the structures 3- and the corresponding radical 2- (panel C on Figure 13 and Figures S31—S33). In all cases, the density difference is located on bases (not phosphate groups). The energy and dynamics of these states may be strongly influenced by the multiply charged character of gas-phase polyanions, and this may explain the changes of the action spectra as a function of the charge state.

## <u>Conclusions</u>

UVMPD action spectroscopy gives a fair reflection of the absorption spectra in the gas phase, but to perform ion spectroscopy on large biomolecules, will suffer from a degree-of-freedom problem: the fragmentation efficiency gets lower as the size of the system increases, and the number of photons required to observe fragmentation increases. When studying multiply charged anions produced by electrospray ionization, another frequently observed action channel is electron photodetachment. It offers the advantage of being mono-photonic, and thus ePD action spectroscopy will remain sensitive when the size of the system increases. However, through the present systematic study, we have questioned here for the first time whether ePD action spectroscopy only reflects the UV absorption spectrum.

First, the charge state (or change density) influences the shape of the ePD action spectrum: as the charge state increases, the efficiency at low photon energy increases. Second, although resonant absorption by electronic excitation may enhance electron photodetachment (which is a pre-requisite for using ePD action spectroscopy to infer electronic absorption properties), the efficiency of the "booster" effect is chromophore-dependent and state-dependent. For example, pyrimidines studied so far do not possess this booster effect in the base absorption band. Excitation of purines boosts the electron detachment, but it is possible that not all states favor ePD with the same efficiency.

Our study in turn opens several intriguing questions, for example: what makes a state "ePD-competent"? The fast nature of ePD will complicate the experimental investigation of the nature of these states, and theoretical insight would be welcome. With regard to applications, an intriguing question is: if an ePD-competent chromophore is introduced in a large system, will this chromophore report for its own environment only, or for all the absorbing states of the entire system? For oligonucleotides, we plan to study this by introducing for example a single purine in a pyridine-rich structure.

In summary, electron photodetachment action spectroscopy reflects something more than just absorption, because its efficiency is boosted by special states in some molecular systems. Furthermore, the electron photodetachment effect observed in the gas phase may have a similar origin as the 1-photon oxidation observed in solution at 266 nm for duplexes[43-45] and for guanine G-quadruplexes.[46, 47] Experiments based on electron photodetachment from DNA multi-anions may thus also serve to shed light on the fundamental processes of DNA photodamage, photo-oxidation, mutations and cancer.

**Acknowledgements**

This work was support by the European Research Council under the European Union's Seventh Framework Program (ERC grant 616551 to VG). VG acknowledges Katerina Matheis and Manfred M. Kappes (Karlsruhe Institute of Technology, Germany) for allowing us to use the photoelectron spectra for the present publication, and for fruitful discussion. The authors also thank the CLIO facility for beam time and technical assistance, and Gilles Grégoire and Joël Lemaire for discussions on IRMPD spectroscopy.

**Conflicts of interest:** there are no conflicts of interest to declare.

## References

1. J. B. Fenn, M. Mann, C. K. Meng and S. F. Wong, *Mass Spectrom. Rev.*, 1990, **9**, 37-70.

2. A. C. Leney and A. J. R. Heck, *J. Am. Soc. Mass Spectrom.*, 2017, **28**, 5-13.

3. B. C. Bohrer, S. I. Merenbloom, S. L. Koeniger, A. E. Hilderbrand and D. E. Clemmer, *Annu. Rev. Anal. Chem.*, 2008, **1**, 293-327.

4. T. Wyttenbach and M. T. Bowers, *Mod. Mass Spectrom.*, 2003, **225**, 207-232.

5. N. C. Polfer and J. Oomens, *Mass Spectrom. Rev.*, 2009, **28**, 468-494.

6. J. L. Mergny, J. Li, L. Lacroix, S. Amrane and J. B. Chaires, *Nucleic Acids Res.*, 2005, **33**, e138.

7. E. Emanuele, D. Markovitsi, P. Millie and K. Zakrzewska, *ChemPhysChem*, 2005, **6**, 1387-1392.

8. M. Huix-Rotllant, J. Brazard, R. Improta, I. Burghardt and D. Markovitsi, *J. Phys. Chem. Lett.*, 2015, **6**, 2247-2251.

9. K. Kleinermanns, D. Nachtigallová and M. S. de Vries, *Int. Rev. Phys. Chem.*, 2013, **32**, 308-342.

10. C. T. Middleton, K. de La Harpe, C. Su, Y. K. Law, C. E. Crespo-Hernandez and B. Kohler, *Annu. Rev. Phys. Chem.*, 2009, **60**, 217-239.

11. D. Markovitsi, *Photochem. Photobiol.*, 2016, **92**, 45-51.

12. V. Gabelica, T. Tabarin, R. Antoine, F. Rosu, I. Compagnon, M. Broyer, E. De Pauw and P. Dugourd, *Anal. Chem.*, 2006, **78**, 6564-6572.

13. V. Gabelica, F. Rosu, T. Tabarin, C. Kinet, R. Antoine, M. Broyer, E. De Pauw and P. Dugourd, *J. Am. Chem. Soc.*, 2007, **129**, 4706-4713.

14. F. Rosu, V. Gabelica, E. De Pauw, R. Antoine, M. Broyer and P. Dugourd, *J. Phys. Chem. A*, 2012, **116**, 5383-5391.

15. V. Gabelica, S. Livet and F. Rosu, *J. Am. Soc. Mass Spectrom.*, 2018, **29**, 2189-2198.

16. A. Marchand, S. Livet, F. Rosu and V. Gabelica, *Anal. Chem.*, 2017, **89**, 12674-12681.

17. K. Arnold, T. S. Balaban, M. N. Blom, O. T. Ehrler, S. Gilb, O. Hampe, J. E. van Lier, J. M. Weber and M. M. Kappes, *J. Phys. Chem. A*, 2003, **107**, 794-803.

18. I. Ivani, P. D. Dans, A. Noy, A. Perez, I. Faustino, A. Hospital, J. Walther, P. Andrio, R. Goni, A. Balaceanu, G. Portella, F. Battistini, J. L. Gelpi, C. Gonzalez, M. Vendruscolo, C. A. Laughton, S. A. Harris, D. A. Case and M. Orozco, *Nat. Methods*, 2016, **13**, 55-58.

19. D. A. Case, T. A. Darden, T. E. Cheatham, C. L. Simmerling, J. Wang, R. E. Duke, R. Luo, R. C. Walker, W. Zhang, K. M. Merz, B. Roberts, S. Hayik, A. Roitberg, G. Seabra, J. Swails, A. W. Goetz, I. Kolossváry, K. F. Wong, F. Paesani, J. Vanicek, R. M. Wolf, J. Liu, X. Wu, S. R. Brozell, T. Steinbrecher, H. Gohlke, Q. Cai, X. Ye, J. Wang, M. J. Hsieh, G. Cui, D. R. Roe, D.

H. Mathews, M. G. Seetin, R. Salomon-Ferrer, C. Sagui, V. Babin, T. Luchko, S. Gusarov, A. Kovalenko and P. A. Kollman, *Journal*, 2012, DOI: citeulike-article-id:10779586.

20.     M. J. Frisch, G. W. Trucks, H. B. Schlegel, G. E. Scuseria, M. A. Robb, J. R. Cheeseman, G. Scalmani, V. Barone, G. A. Petersson, H. Nakatsuji, X. Li, M. Caricato, A. V. Marenich, J. Bloino, B. G. Janesko, R. Gomperts, B. Mennucci, H. P. Hratchian, J. V. Ortiz, A. F. Izmaylov, J. L. Sonnenberg, Williams, F. Ding, F. Lipparini, F. Egidi, J. Goings, B. Peng, A. Petrone, T. Henderson, D. Ranasinghe, V. G. Zakrzewski, J. Gao, N. Rega, G. Zheng, W. Liang, M. Hada, M. Ehara, K. Toyota, R. Fukuda, J. Hasegawa, M. Ishida, T. Nakajima, Y. Honda, O. Kitao, H. Nakai, T. Vreven, K. Throssell, J. A. Montgomery Jr., J. E. Peralta, F. Ogliaro, M. J. Bearpark, J. J. Heyd, E. N. Brothers, K. N. Kudin, V. N. Staroverov, T. A. Keith, R. Kobayashi, J. Normand, K. Raghavachari, A. P. Rendell, J. C. Burant, S. S. Iyengar, J. Tomasi, M. Cossi, J. M. Millam, M. Klene, C. Adamo, R. Cammi, J. W. Ochterski, R. L. Martin, K. Morokuma, O. Farkas, J. B. Foresman and D. J. Fox, *Gaussian 16 Rev. B.01*, Wallingford, CT, 2016.

21.     Y. Zhao and D. G. Truhlar, *Theor. Chem. Acc.*, 2008, **120**, 215-241.

22.     T. A. Zubatiuk, O. V. Shishkin, L. Gorb, D. M. Hovorun and J. Leszczynski, *Phys. Chem. Chem. Phys.*, 2013, **15**, 18155-18166.

23.     M. F. Mesleh, J. M. Hunter, A. A. Shvartsburg, G. C. Schatz and M. F. Jarrold, *J. Phys. Chem.*, 1996, **100**, 16082-16086.

24.     J. Wu and S. A. McLuckey, *Int. J. Mass Spectrom.*, 2004, **237**, 197-241.

25.     R. Cercola, E. Matthews and C. E. H. Dessent, *J. Phys. Chem. B*, 2017, **121**, 5553-5561.

26.     L. Joly, R. Antoine, M. Broyer, J. Lemoine and P. Dugourd, *J. Phys. Chem. A*, 2008, **112**, 898-903.

27.     C. Brunet, R. Antoine, J. Lemoine and P. Dugourd, *J. Phys. Chem. Lett.*, 2012, **3**, 698-702.

28.     Y. Liu, L. Martínez-Fernández, J. Cerezo, G. Prampolini, R. Improta and F. Santoro, *Chem. Phys.*, 2018, **in press**, 10.1016/j.chemphys.2018.1008.1030.

29.     R. Improta, F. Santoro and L. Blancafort, *Chem. Rev.*, 2016, **116**, 3540-3593.

30.     K. Matheis, Photoelektronenspektroskopie an mehrfach geladenen Anionen in der Gasphase, PhD dissertation, Karlsruher Institut für Technologie (KIT) − Universitätsbereich, 2010.

31.     J. M. Weber, I. N. Ioffe, K. M. Berndt, D. Loffler, J. Friedrich, O. T. Ehrler, A. S. Danell, J. H. Parks and M. M. Kappes, *J. Am. Chem. Soc.*, 2004, **126**, 8585-8589.

32.     M. Vonderach, O. T. Ehrler, K. Matheis, P. Weis and M. M. Kappes, *J. Am. Chem. Soc.*, 2012, **134**, 7830-7841.

33.     L. M. Nielsen, S. O. Pedersen, M. B. Kirketerp and S. B. Nielsen, *J. Chem. Phys.*, 2012, **136**, 064302.

34.     J. S. Klassen, P. D. Schnier and E. R. Williams, *J. Am. Soc. Mass Spectrom.*, 1998, **9**, 1117-1124.

35.     D. Markovitsi, T. Gustavsson and I. Vayá, *J. Phys. Chem. Lett.*, 2010, **1**, 3271-3276.

36.     R. Antoine and P. Dugourd, *Phys. Chem. Chem. Phys.*, 2011, **13**, 16494-16509.

37.     M. N. Ashfold, G. A. King, D. Murdock, M. G. Nix, T. A. Oliver and A. G. Sage, *Phys. Chem. Chem. Phys.*, 2010, **12**, 1218-1238.

38.     S. Yamazaki, W. Domcke and A. L. Sobolewski, *J. Phys. Chem. A*, 2008.

39.     T. N. V. Karsili, B. Marchetti and M. N. R. Ashfold, *Chem. Phys.* , 2018, **in press**, 10.1016/j.chemphys.2018.1008.1016.

40.     H. H. Ritze, H. Lippert, E. Samoylova, V. R. Smith, I. V. Hertel, W. Radloff and T. Schultz, *J. Chem. Phys.*, 2005, **122**, 224320.

41.     V. A. Spata, W. Lee and S. Matsika, *J. Phys. Chem. Lett.*, 2016, **7**, 976-984.

42.     G. Olaso-Gonzalez, M. Merchan and L. Serrano-Andres, *J. Am. Chem. Soc.*, 2009, **131**, 4368-4377.

43.     S. Marguet, D. Markovitsi and F. Talbot, *J. Phys. Chem. B*, 2006, **110**, 11037-11039.

44.     A. Banyasz, T. Ketola, L. Martinez-Fernandez, R. Improta and D. Markovitsi, *Faraday Discuss.*, 2018, **207**, 181-197.

45.     A. Banyasz, L. Martinez-Fernandez, R. Improta, T. M. Ketola, C. Balty and D. Markovitsi, *Phys. Chem. Chem. Phys.*, 2018, **20**, 21381-21389.

46.     L. Martinez-Fernandez, A. Banyasz, D. Markovitsi and R. Improta, *Chem. Eur. J.*, 2018, **24**, 15185-15189.

47.     A. Banyasz, L. Martinez-Fernandez, C. Balty, M. Perron, T. Douki, R. Improta and D. Markovitsi, *J. Am. Chem. Soc.*, 2017, **139**, 10561-10568.

**SUPPORTING INFORMATION: supplementary figures as described in the text**

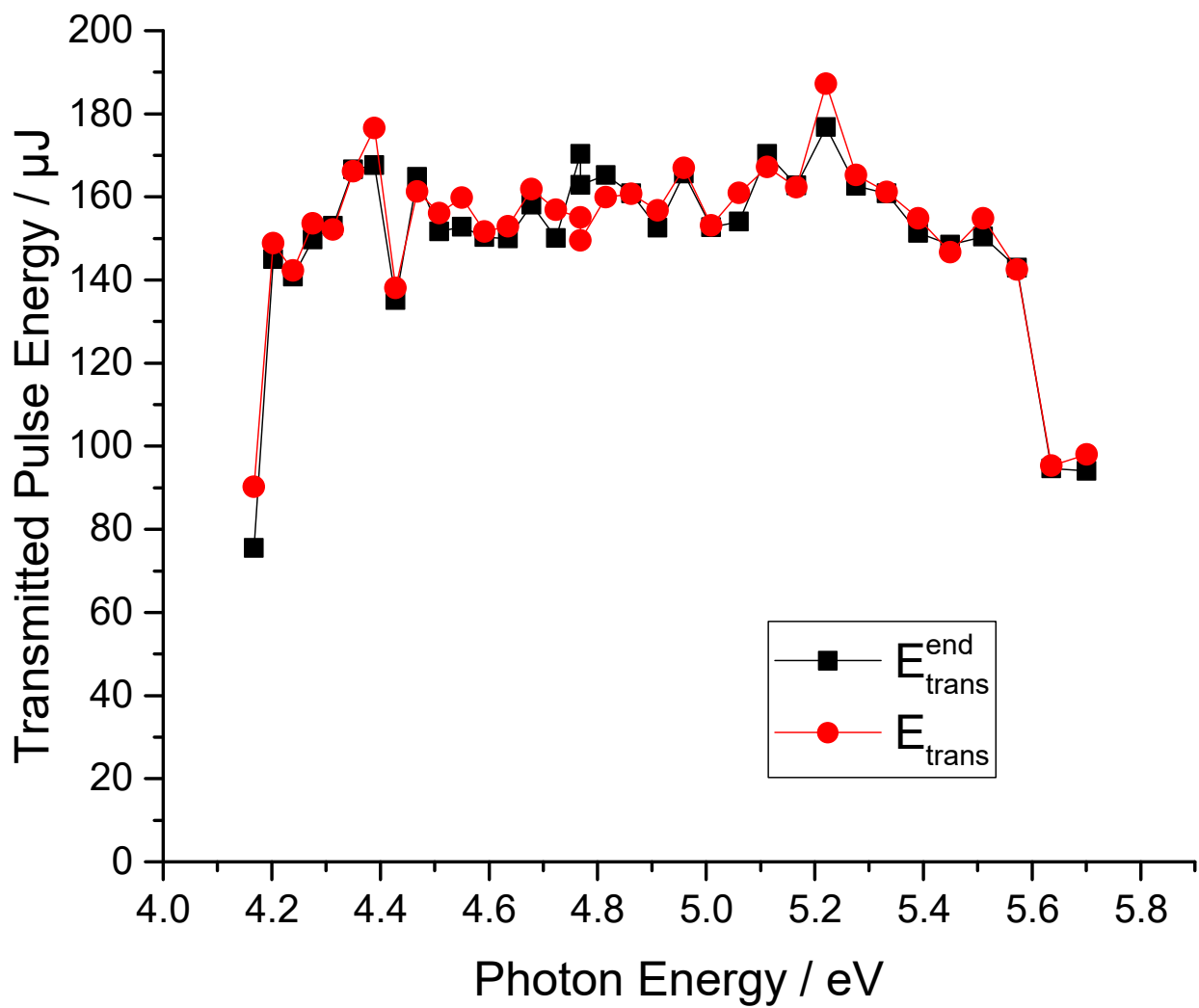

Figure S1. Representative data for the transmitted pulse energy as a function of the photon energy. The black points were measured immediately following acquisition of a mass spectrum, concurrently with measurement of the reflected pulse energy. The red points are calculated pulse energy used to irradiate ions during acquisition of a mass spectrum (see main text). The extreme photon energy points have lower pulse energy due to reaching the edge of the tuning range of the OPO.

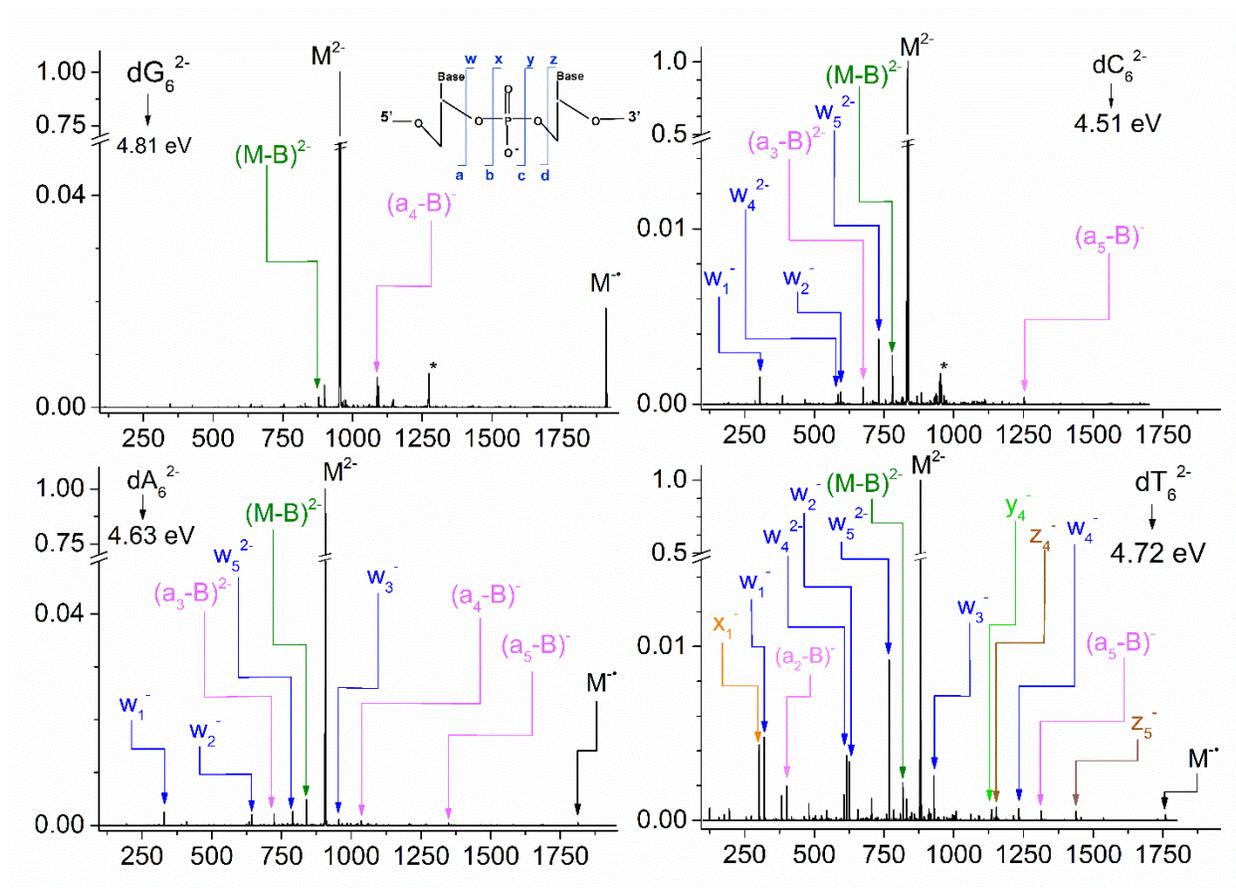

Figure S2. Mass spectra following irradiation with a single UV laser pulse (160 µJ transmitted through the trap) at 4.81 eV (257.5 nm) for $dG_6^{2-}$ (top left), 4.51 eV (275 nm) for $dC_6^{2-}$ (top right), 4.63 eV (267.5 nm) nm for $dA_6^{2-}$ (bottom left) and 4.72 eV (262.5 nm) for $dT_6^{2-}$ (bottom right). Major fragmentation channels are labelled according to the nomenclature in the inset, and EPD products are shown in red. Note that the vertical scale for purines (A and G) differs from that of the pyrimidines (C and T). The asterisks denote fragmentation channels come from 4- dimer (which is isobaric). These fragments are excluded from the analysis.

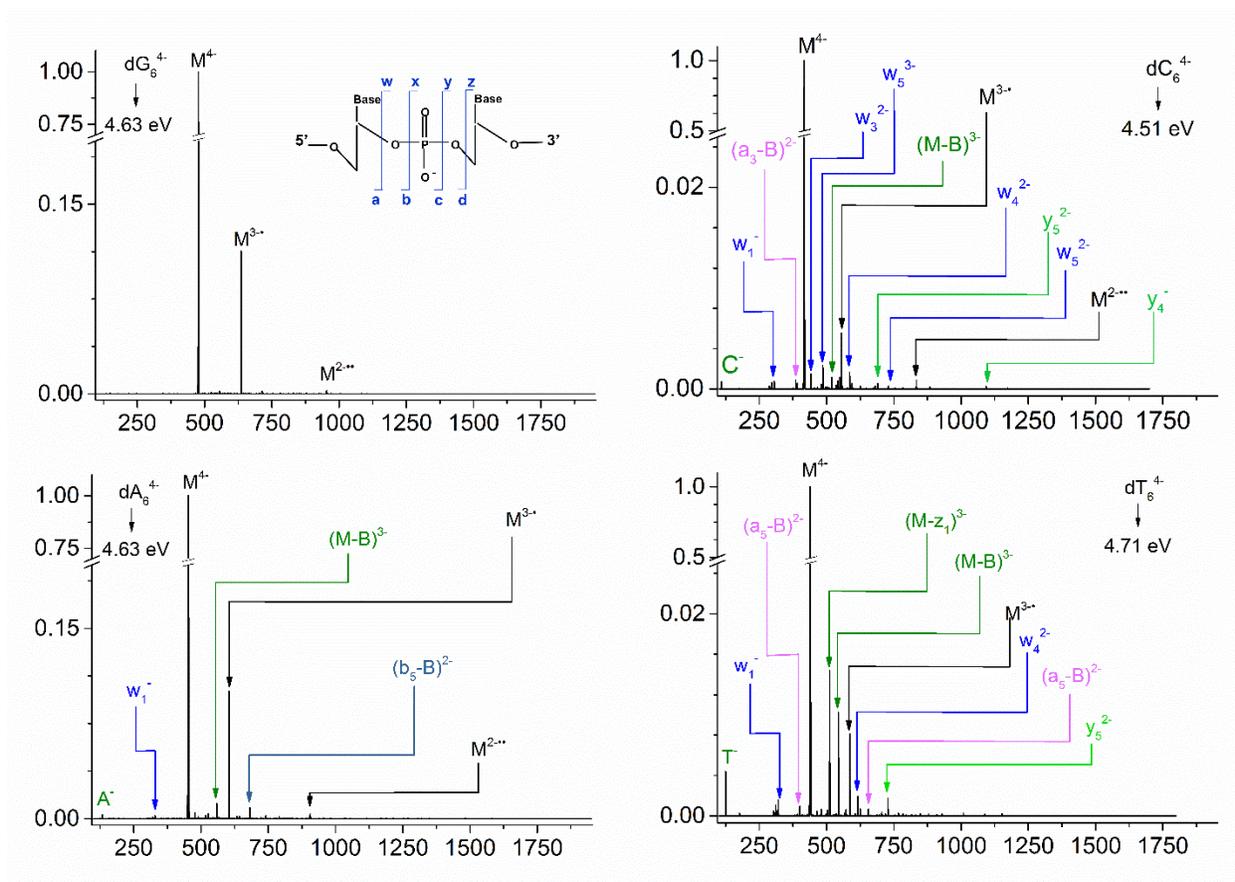

Figure S3. Mass spectra following irradiation with a single UV laser pulse (160 µJ transmitted through the trap) at 4.81 eV (257.5 nm) for $dG_6^{4-}$ (top left), 4.51 eV (275 nm) for $dC_6^{4-}$ (top right), 4.63 eV (267.5 nm) nm) for $dA_6^{4-}$ (bottom left) and 4.72 eV (262.5 nm) for $dT_6^{4-}$ (bottom right). Major fragmentation channels are labelled according to the nomenclature in the inset, and EPD products are shown in red. Note that the vertical scale for purines (A and G) differs from that of the pyrimidines (C and T).

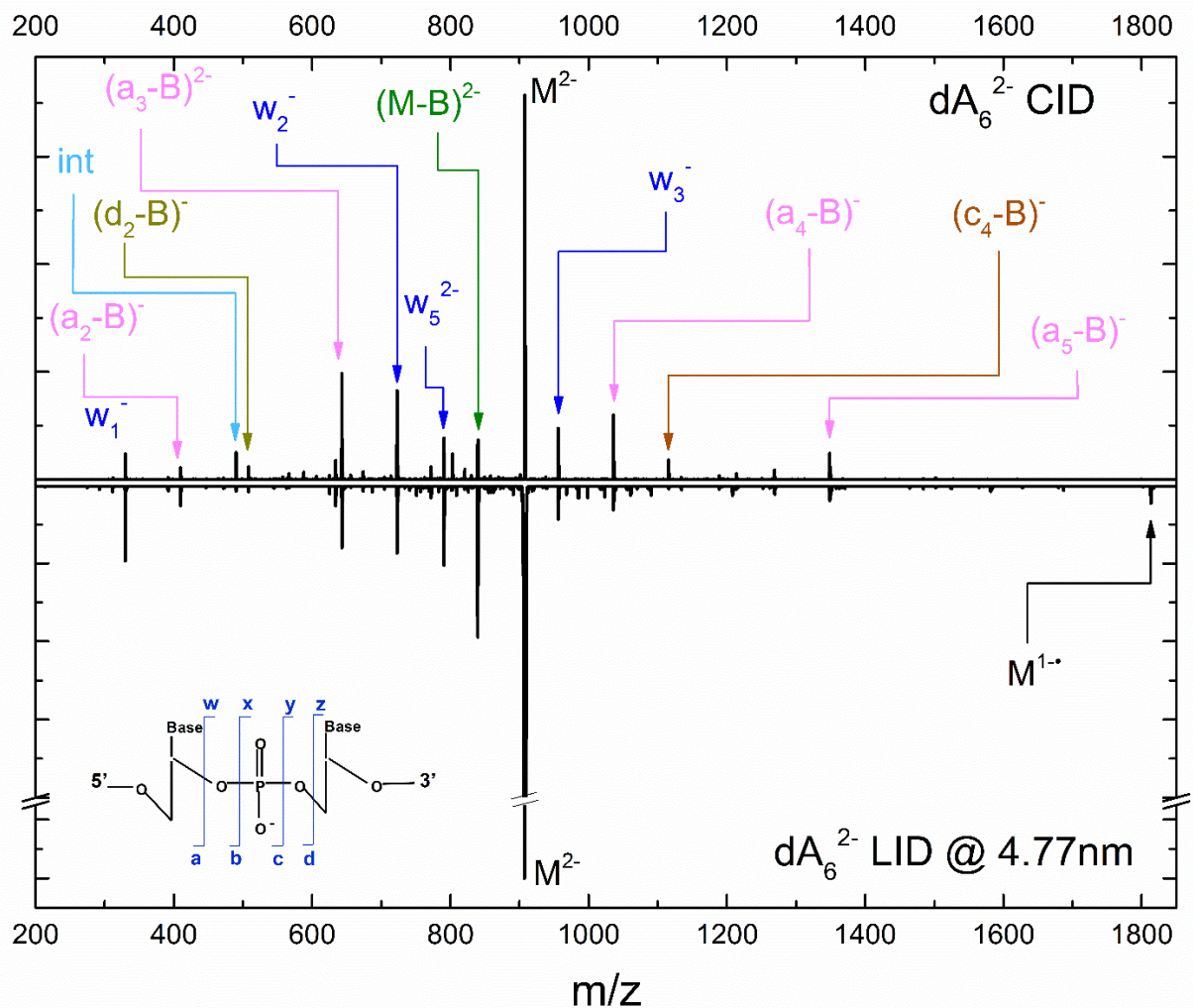

Figure S4. Mass spectra following 10ms collision induced dissociation with an activation voltage of 0.67V (top) and irradiation with 4.77 eV photons (bottom) for $A_6^{2-}$. The major peaks are annotated according to the scheme in the bottom panel. Here int refers to an internal fragment.

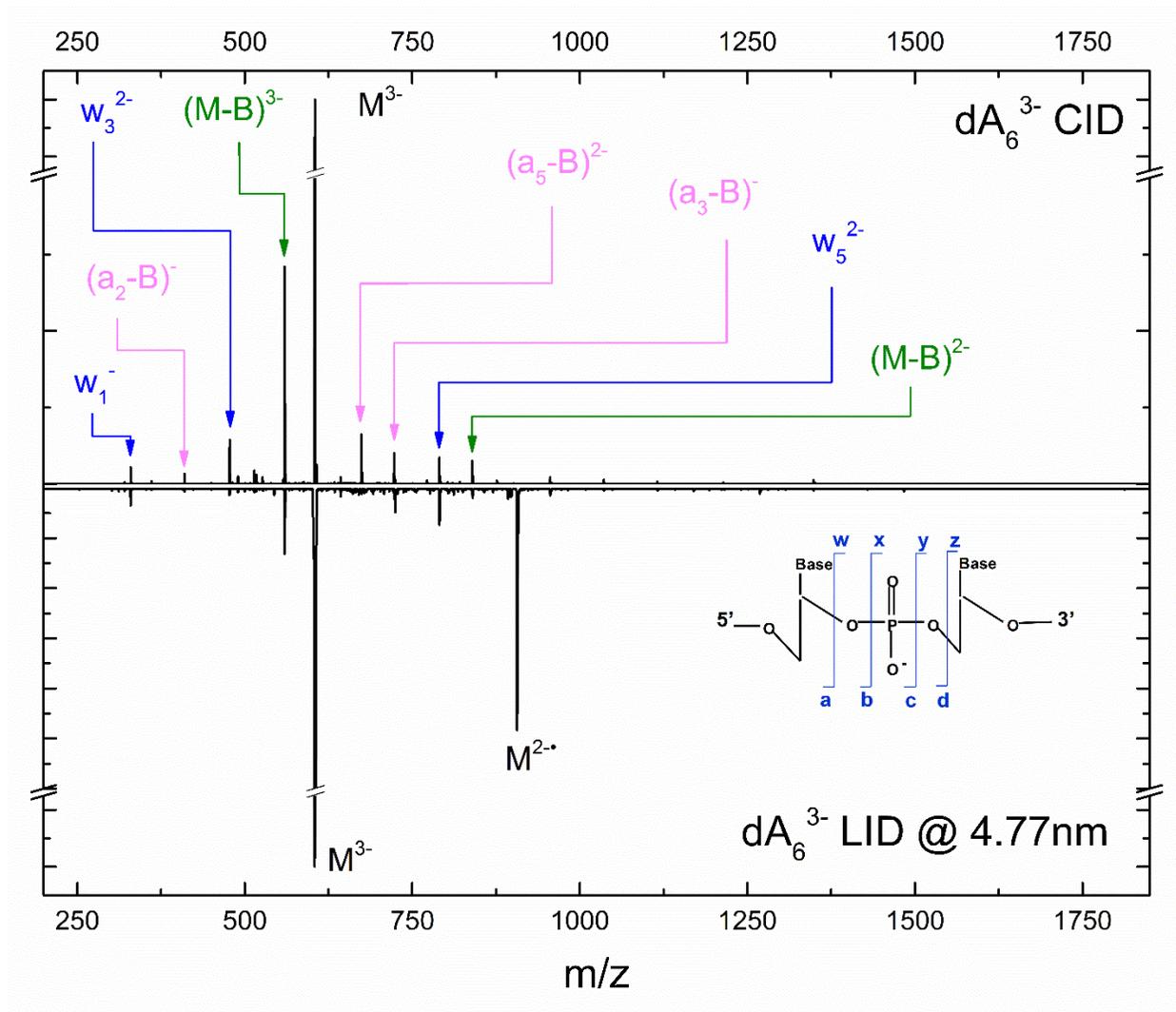

Figure S5. Mass spectra following 10ms collision induced dissociation with an activation voltage of 0.70V (top) and irradiation with 4.77 eV photons (bottom) for $A_6^{3-}$. The major peaks are annotated according to the scheme in the bottom panel.

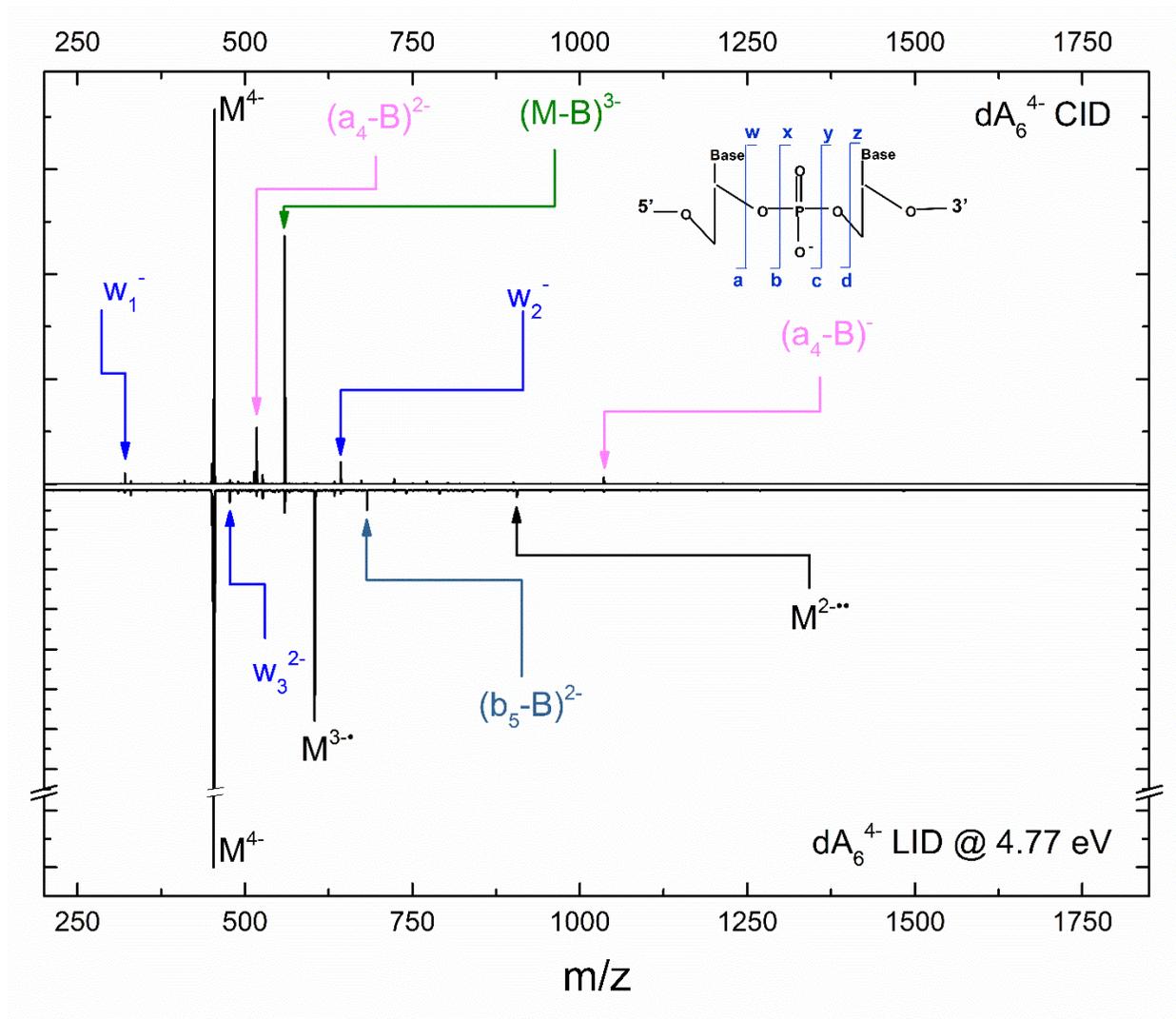

Figure S6. Mass spectra following 10ms collision induced dissociation with an activation voltage of 0.70V (top) and irradiation with 4.77 eV photons (bottom) for $A_6^{4-}$. The major peaks are annotated according to the scheme in the top panel.

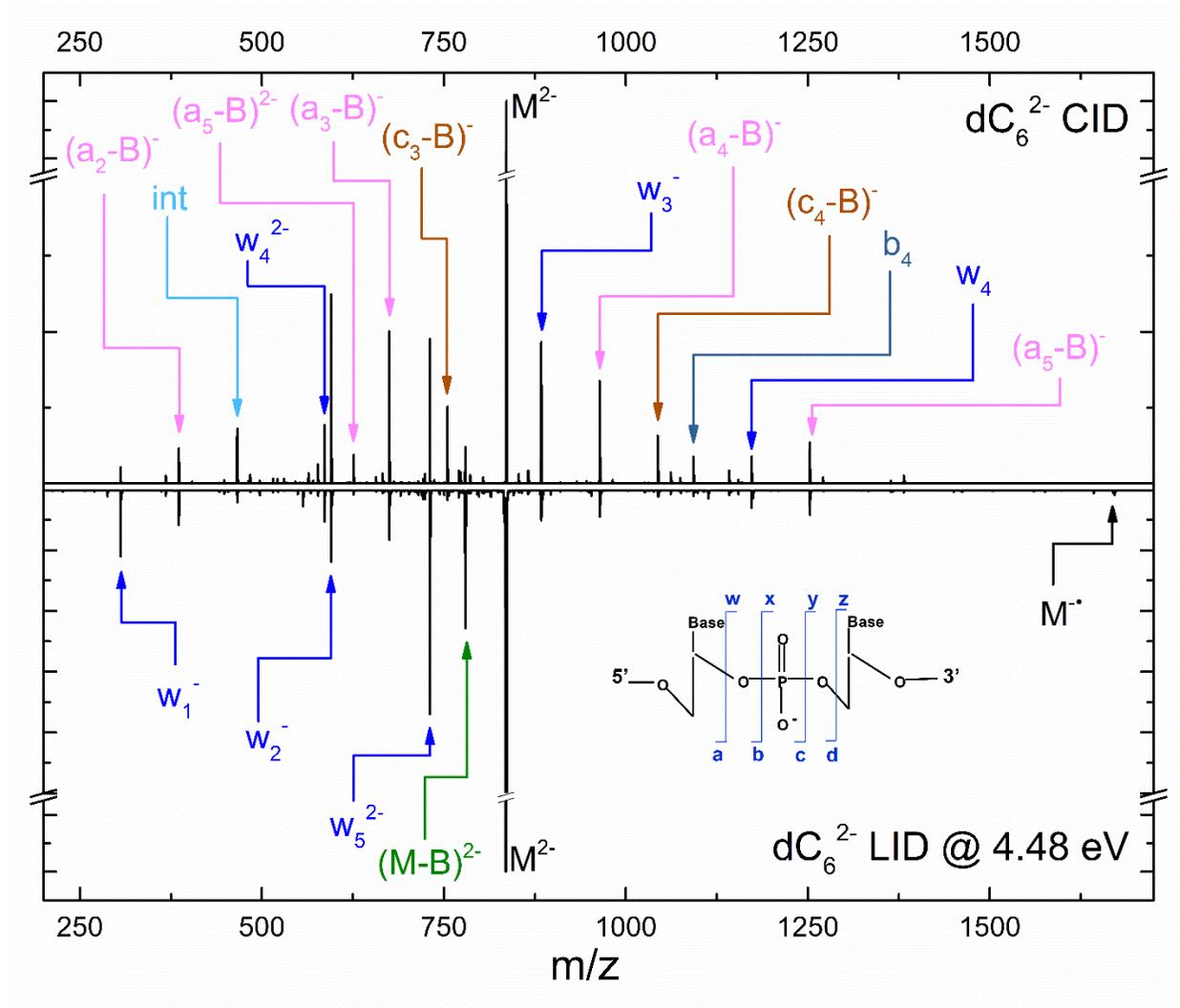

Figure S7. Mass spectra following 10ms collision induced dissociation with an activation voltage of 0.70V (top) and irradiation with 4.48 eV photons (bottom) for $C_6^{2-}$. The major peaks are annotated according to the scheme in the bottom panel.

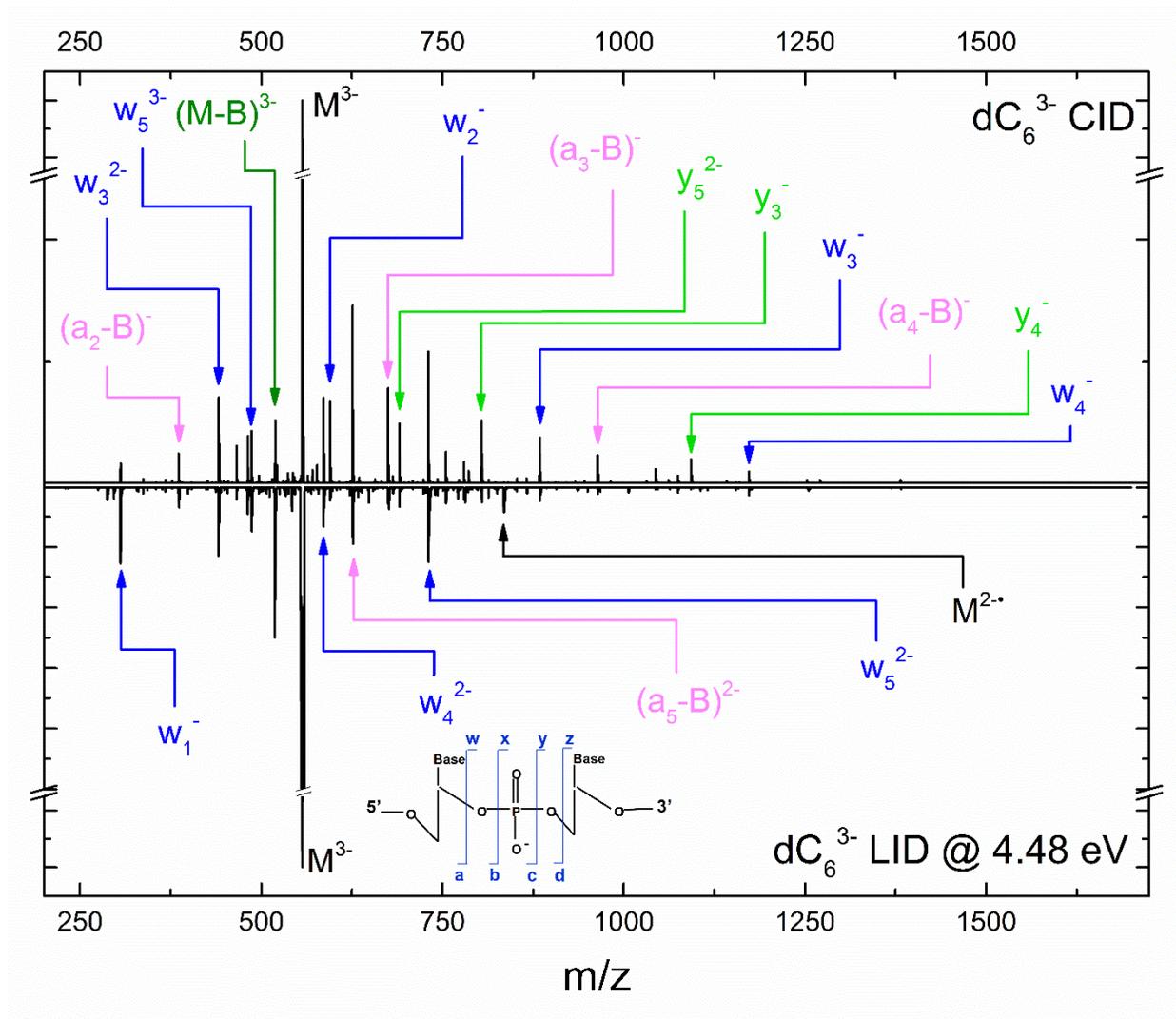

Figure S8. Mass spectra following 10ms collision induced dissociation with an activation voltage of 0.70V (top) and irradiation with 4.48 eV photons (bottom) for $C_6^{3-}$. The major peaks are annotated according to the scheme in the bottom panel.

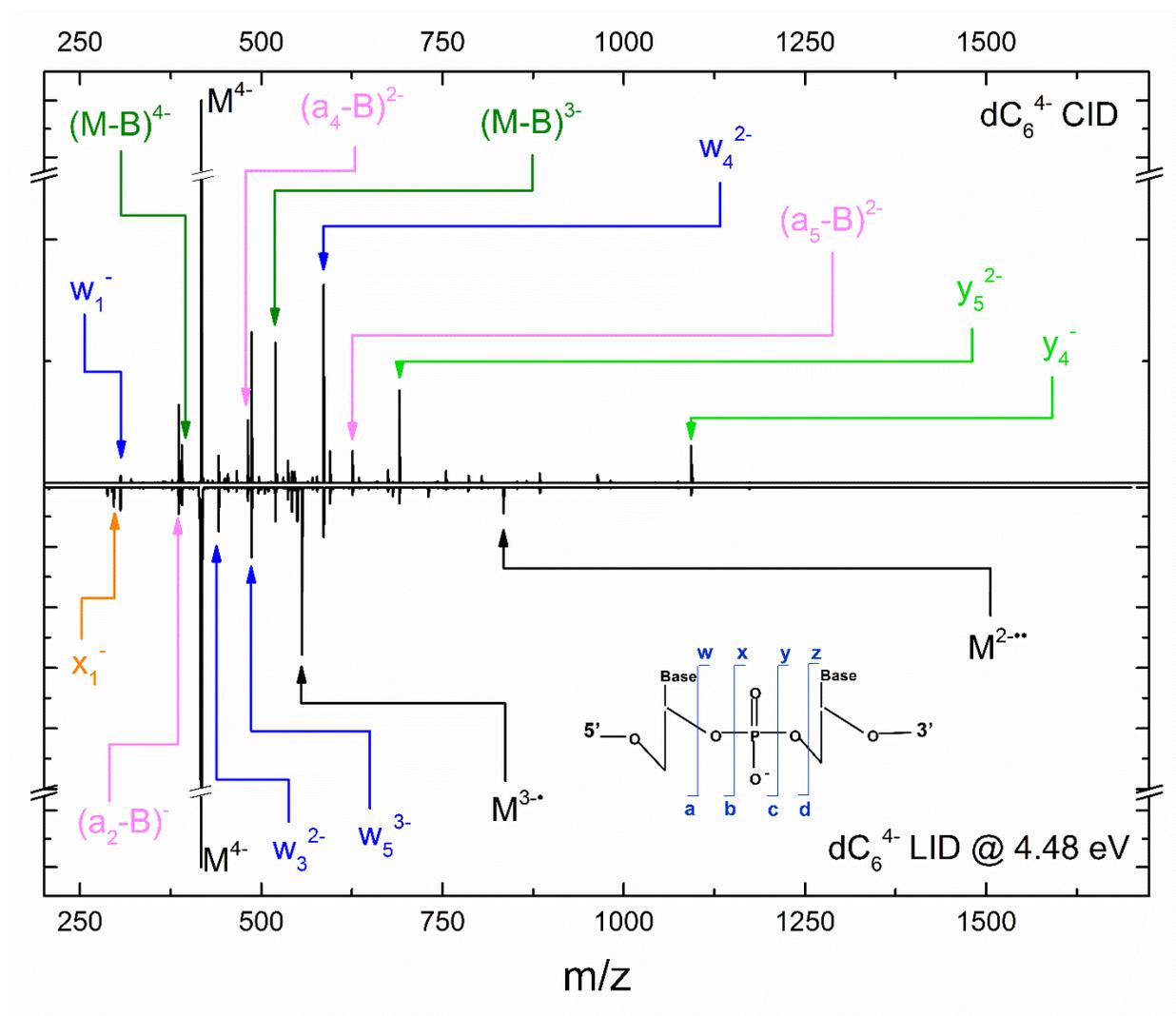

Figure S9. Mass spectra following 10ms collision induced dissociation with an activation voltage of 0.9V (top) and irradiation with 4.48 eV photons (bottom) for $C_6^{4-}$. The major peaks are annotated according to the scheme in the bottom panel.

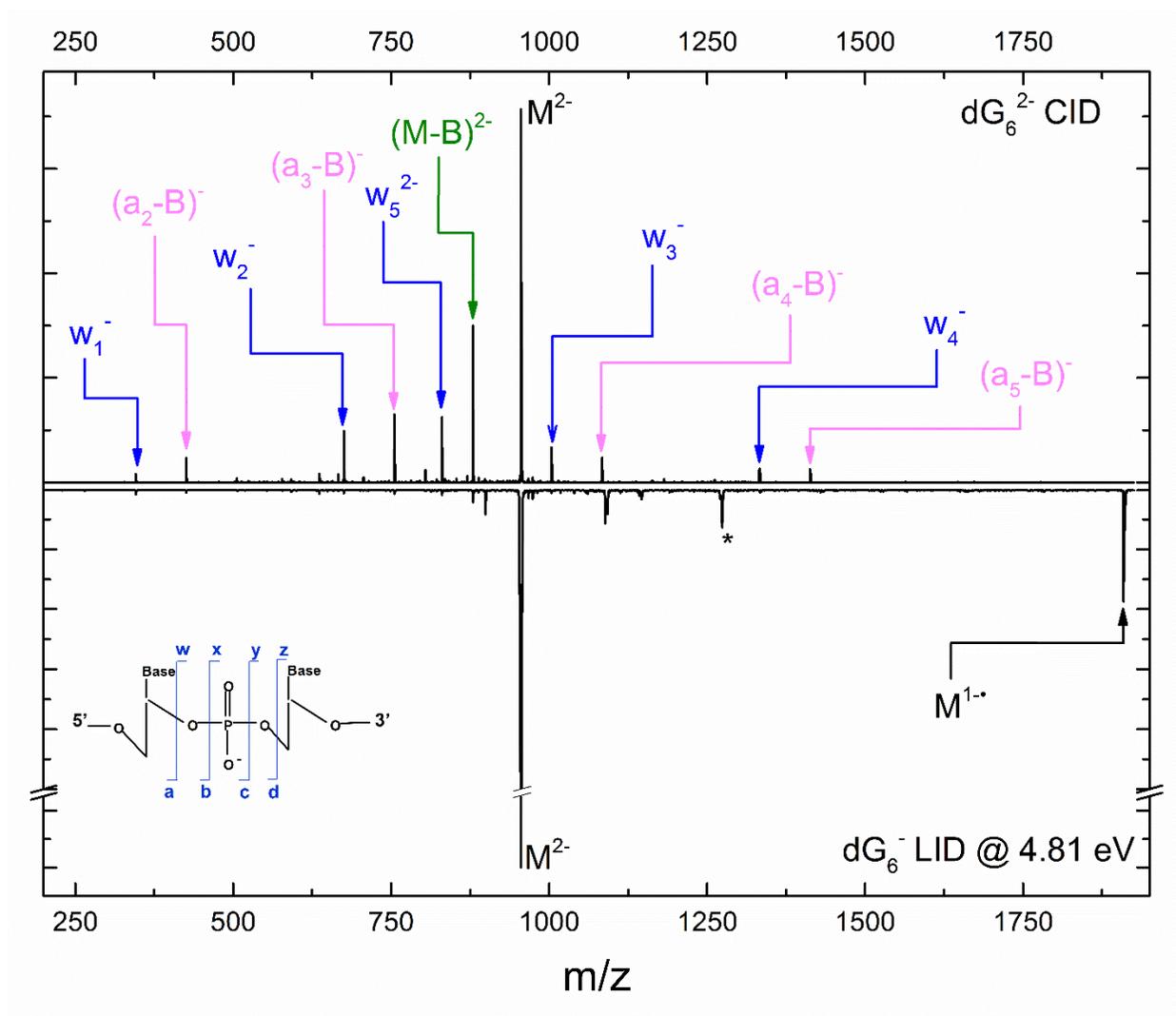

Figure S10. Mass spectra following 10ms collision induced dissociation with an activation voltage of 0.47V (top) and irradiation with 4.81 eV photons (bottom) for $G_6^{2-}$. The major peaks are annotated according to the scheme in the bottom panel.

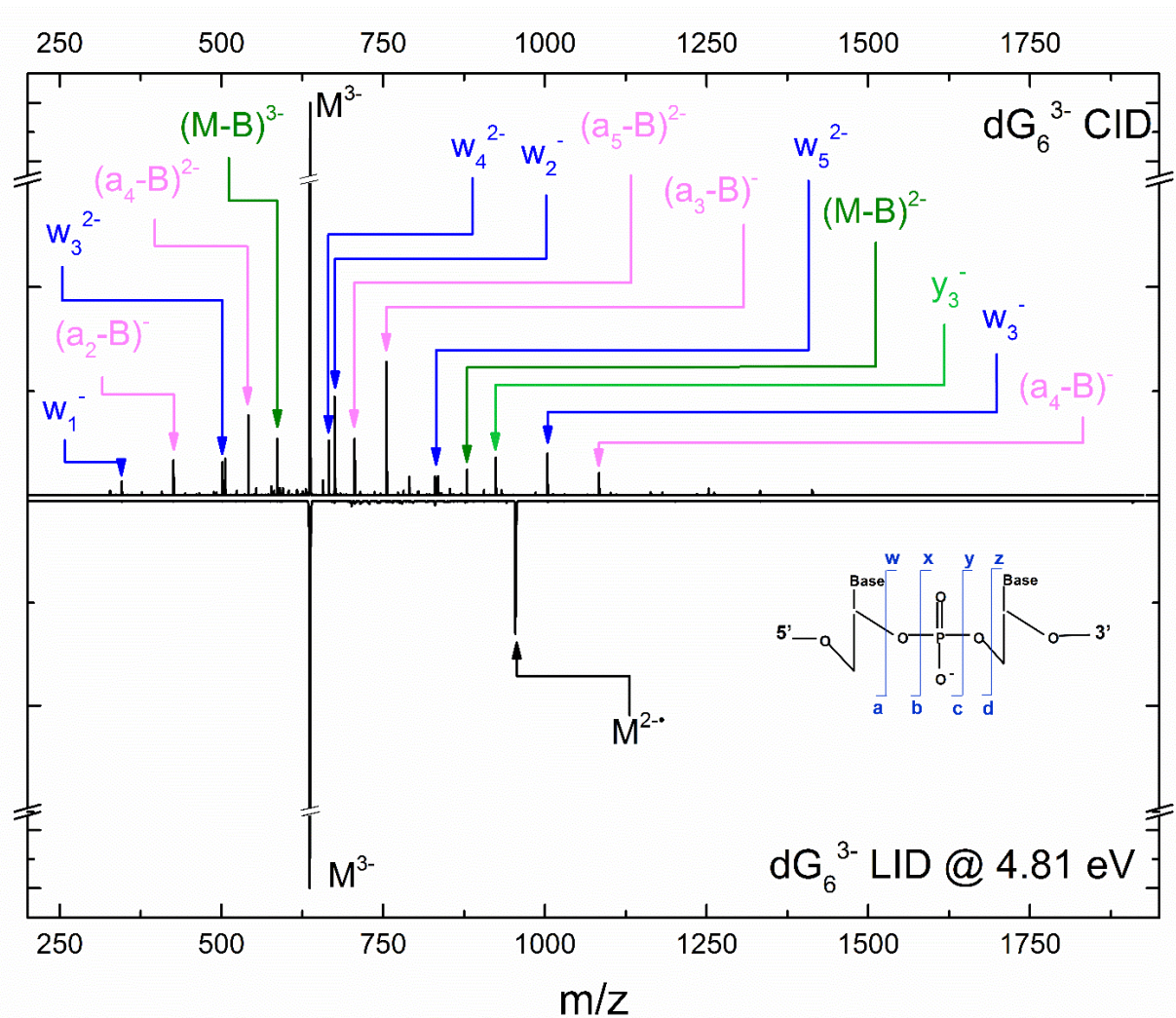

Figure S11. Mass spectra following 10ms collision induced dissociation with an activation voltage of 0.80V (top) and irradiation with 4.81 eV photons (bottom) for $G_6^{3-}$. The major peaks are annotated according to the scheme in the bottom panel.

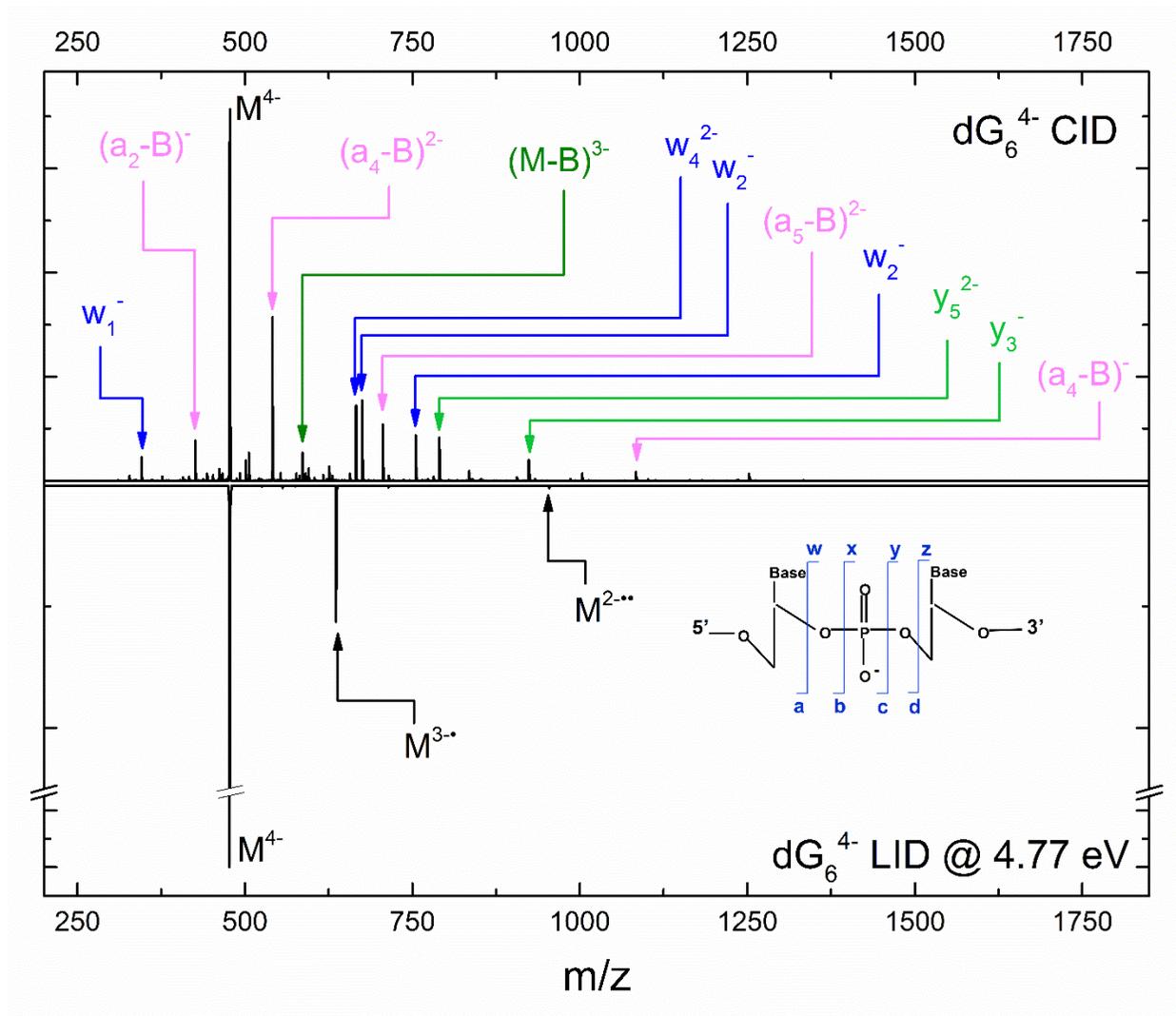

Figure S12. Mass spectra following 10ms collision induced dissociation with an activation voltage of 0.90V (top) and irradiation with 4.81 eV photons (bottom) for $G_6{}^{4-}$. The major peaks are annotated according to the scheme in the bottom panel.

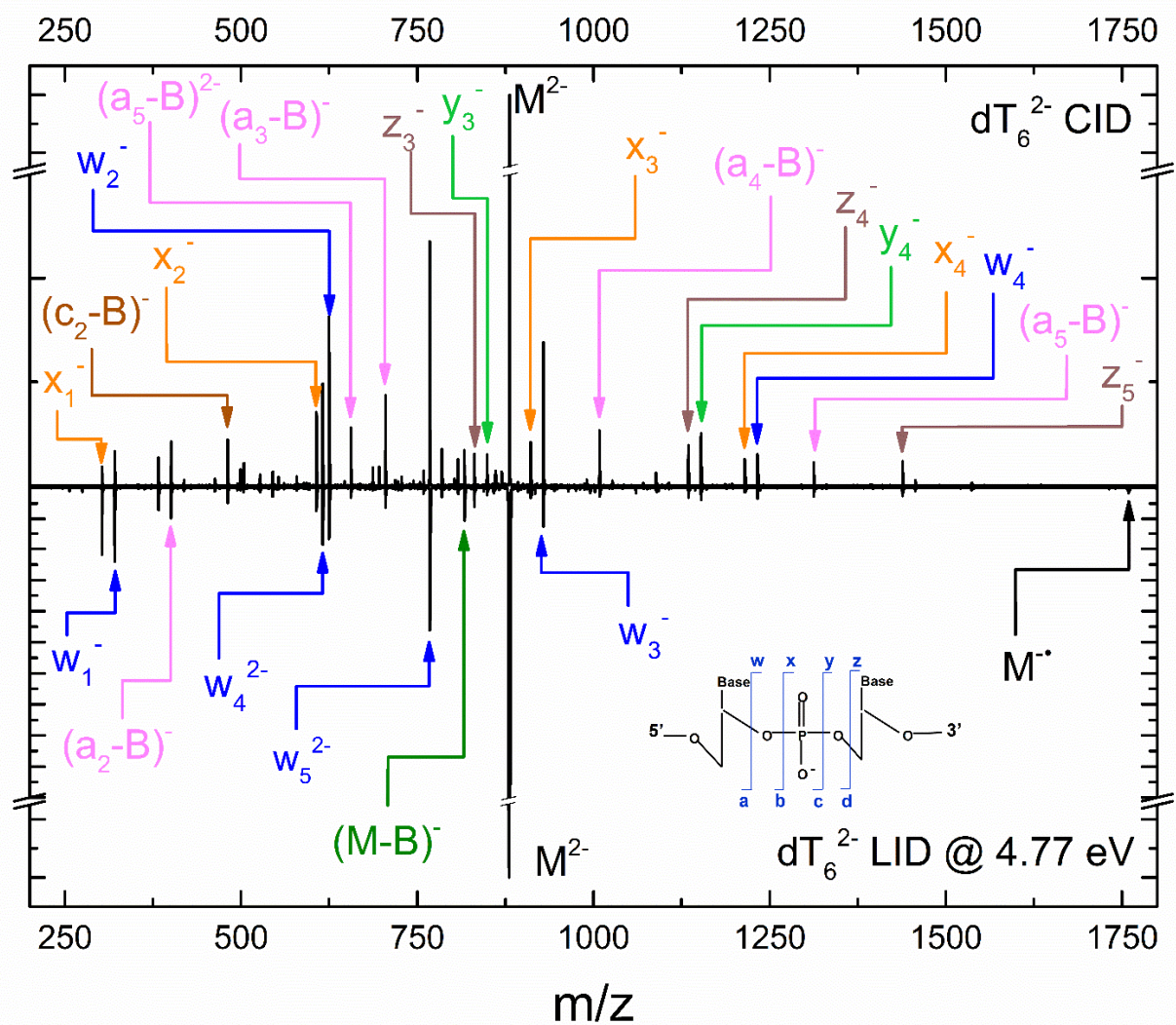

Figure S13. Mass spectra following 10ms collision induced dissociation with an activation voltage of 0.73V (top) and irradiation with 4.77 eV photons (bottom) for $T_6^{2-}$. The major peaks are annotated according to the scheme in the bottom panel.

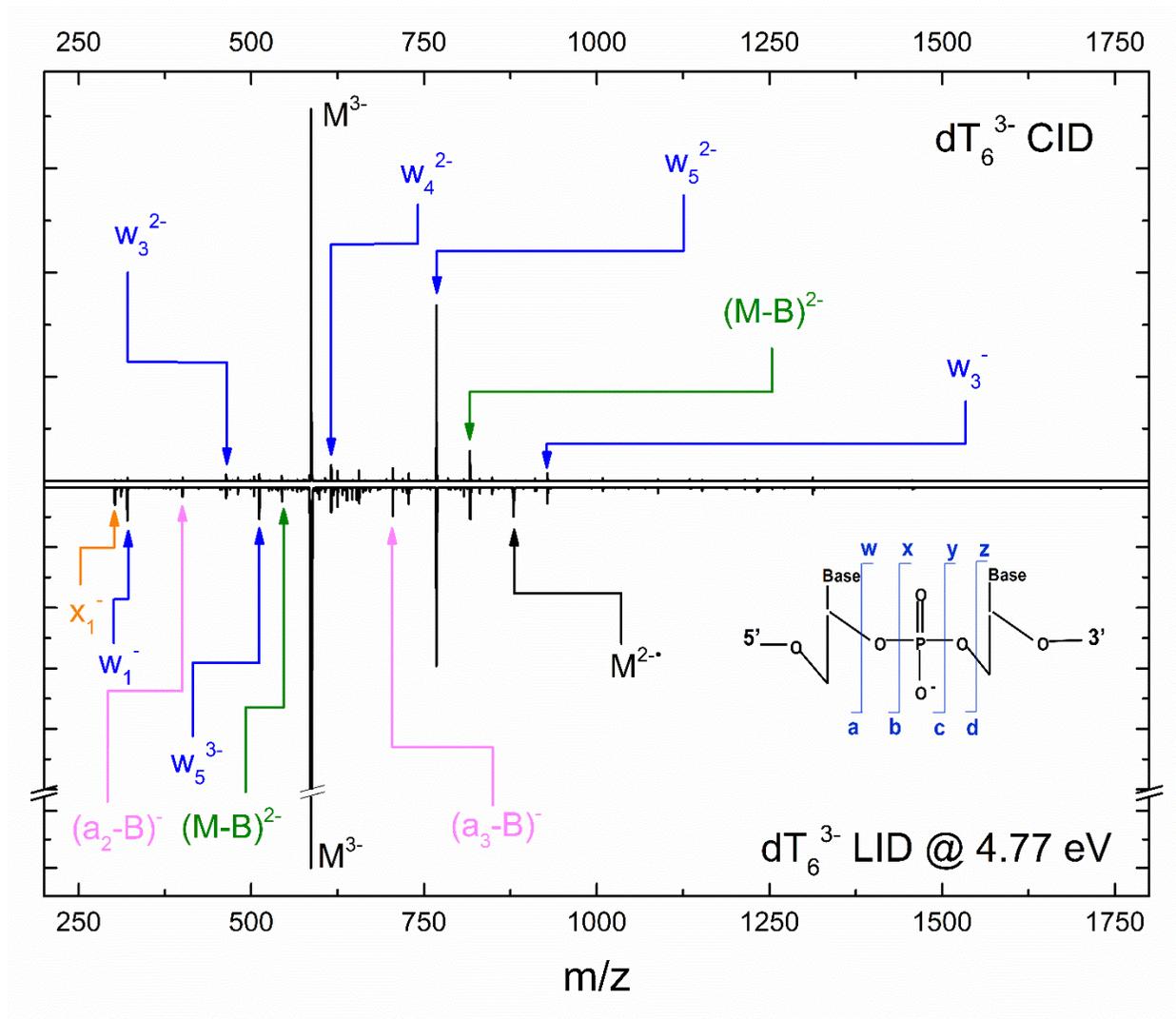

Figure S14. Mass spectra following 10ms collision induced dissociation with an activation voltage of 0.70V (top) and irradiation with 4.77 eV photons (bottom) for $T_6{}^{3-}$. The major peaks are annotated according to the scheme in the bottom panel.

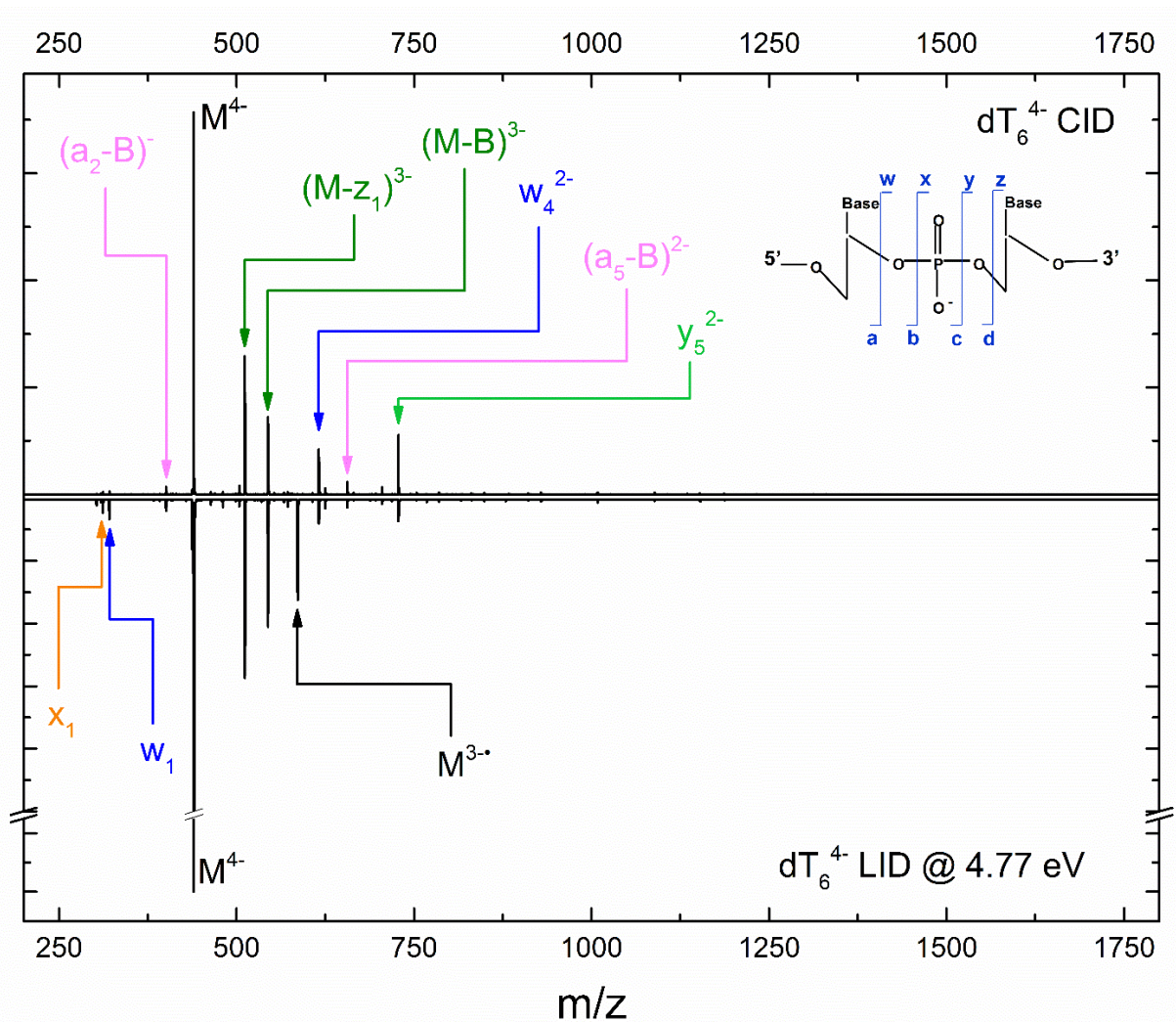

Figure S15. Mass spectra following 10ms collision induced dissociation with an activation voltage of 0.7V (top) and irradiation with 4.77 eV photons (bottom) for $T_6^{4-}$. The major peaks are annotated according to the scheme in the top panel.

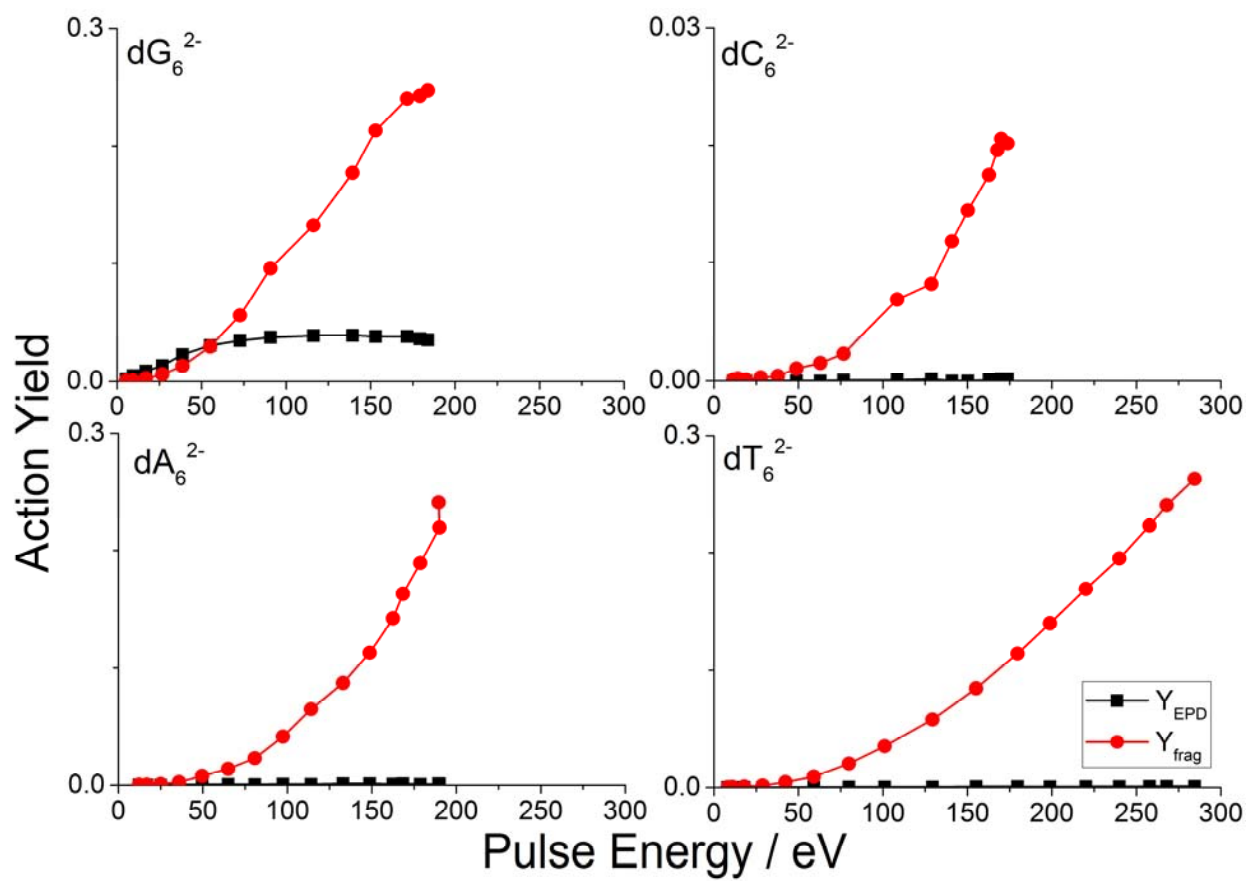

Figure S16. Photoreaction yields versus transmitted pulse energy $E_{trans}$ for EPD (black) and PF (red) for the 2- charge states of dG$_6$ (top left, 4.81 eV), dC$_6$ (top right, 4.51 eV), dA$_6$ (bottom left, 4.63 eV) and dT$_6$ (bottom right, 4.72 eV).

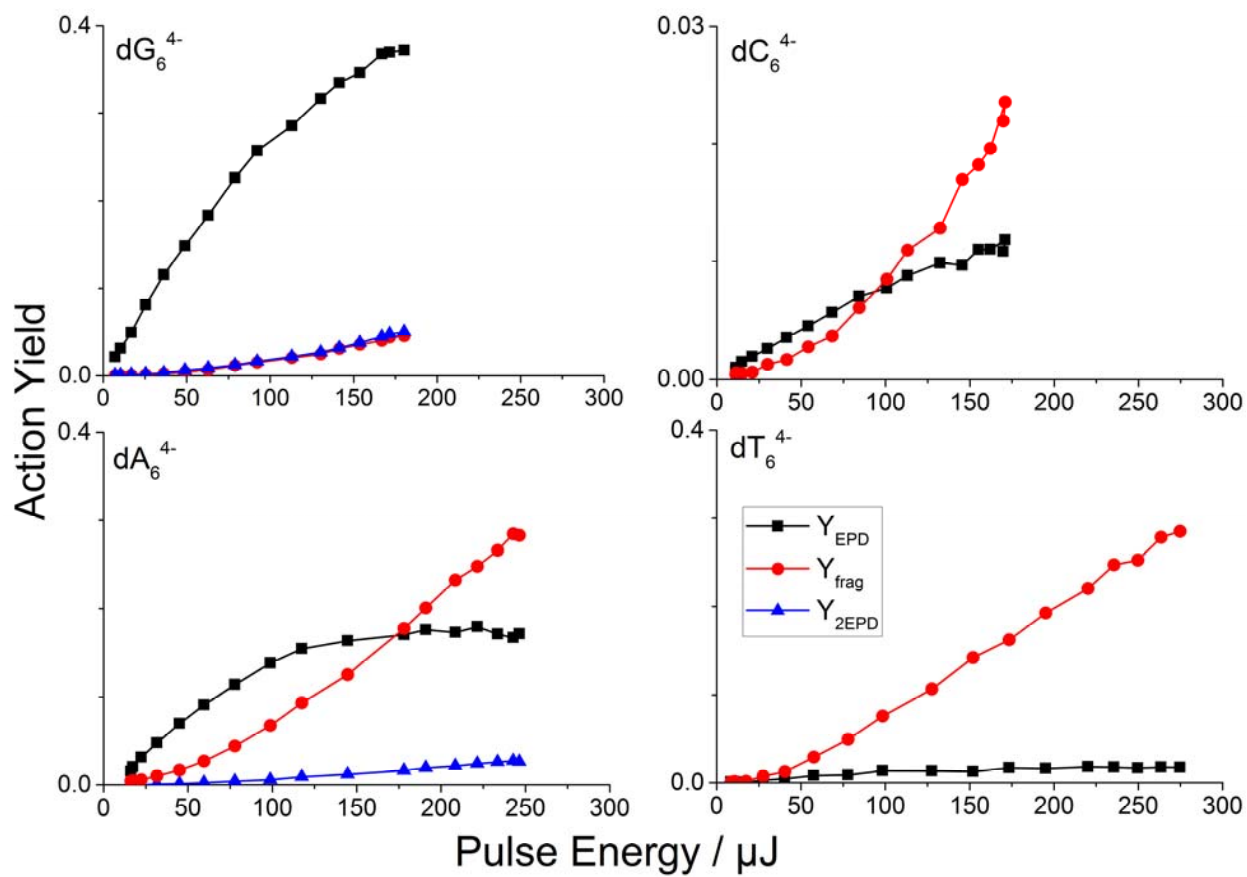

Figure S17. Photoreaction yields versus transmitted pulse energy $E_{trans}$ for EPD (black), PF (red, and loss of two electrons (2EPD) (blue) for the 4- charge states of $dG_6$ (top left, 4.81 eV), $dC_6$ (top right, 4.51 eV), $dA_6$ (bottom left, 4.63 eV) and $dT_6$ (bottom right, 4.72 eV). The dependence of loss of two electrons is quadratic, indicating two sequential electron losses: $M^{4-} \rightarrow M^{3-\bullet} \rightarrow M^{2-\bullet\bullet}$.

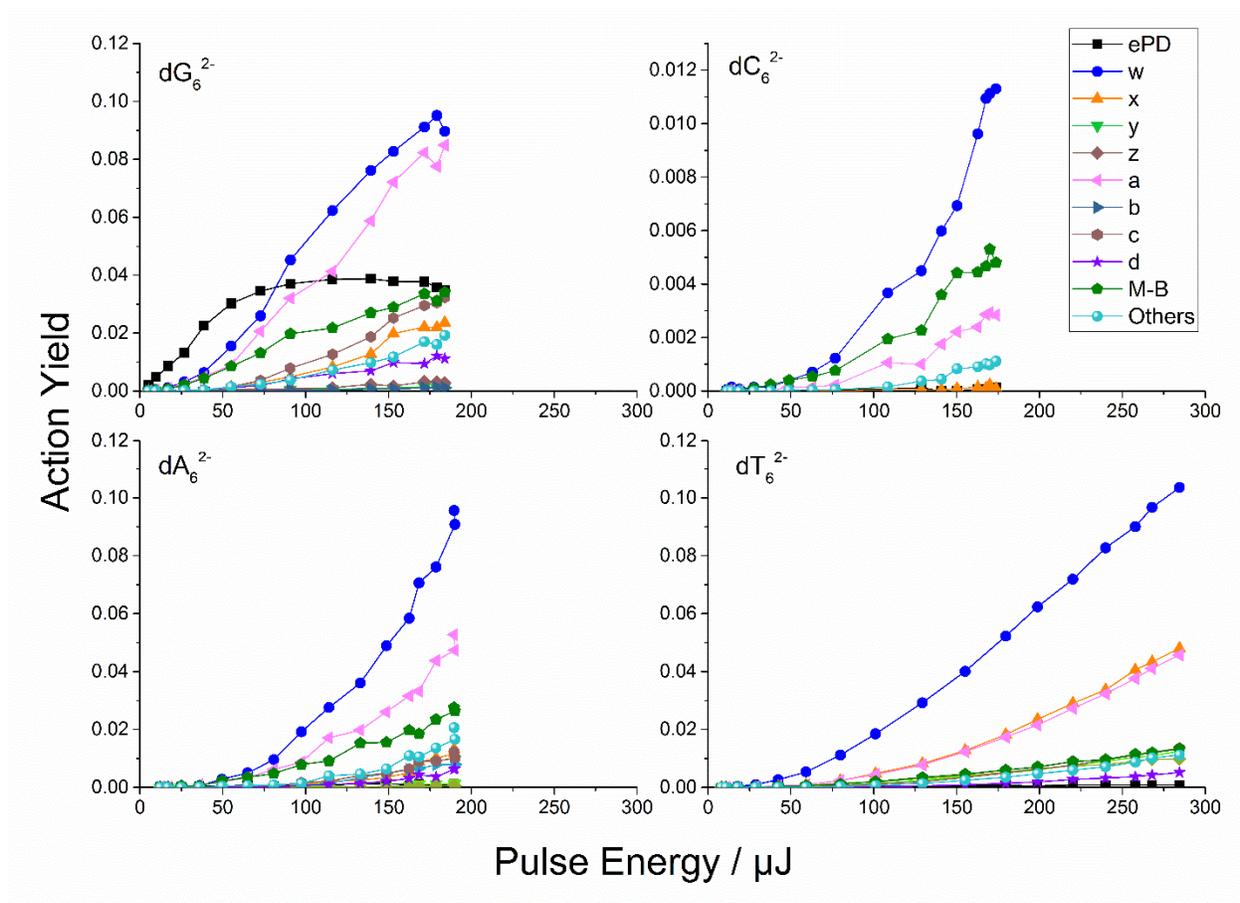

Figure S18. Action yield versus pulse energy for fragments assigned to different families for the 2- charge states of $G_6$ (top left), $C_6$(top right), $A_6$ (bottom left) and $T_6$ (top right). Note the choice of colour reflects the colours used in the annotated MS. None of these are linear, and no individual component is either.

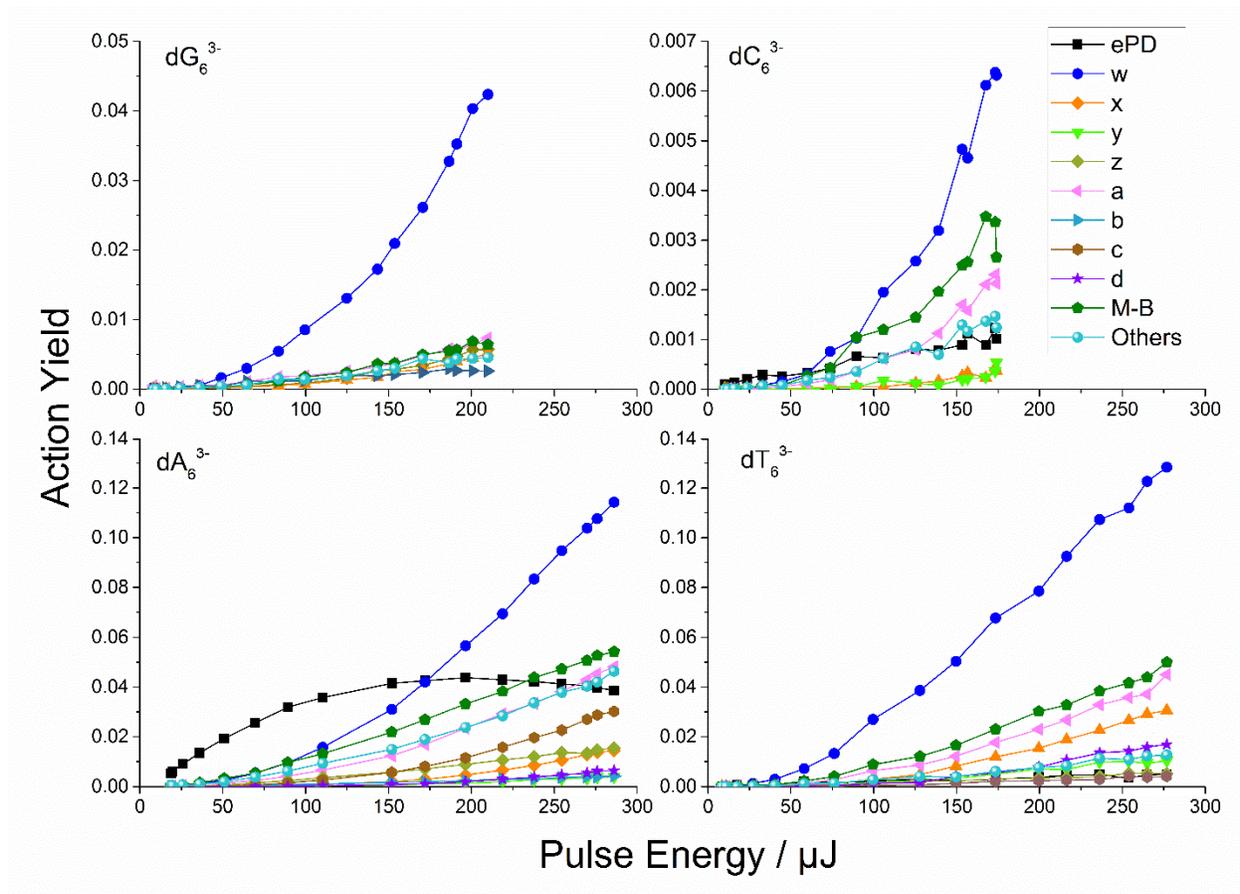

Figure S19. Action yield versus pulse energy for fragments assigned to different families for the 3- charge states of $G_6$ (top left), $C_6$(top right), $A_6$ (bottom left) and $T_6$ (top right). Note the choice of colour reflects the colours used in the annotated MS. None of these are linear, and no individual component is either. The ePD is omitted from the $G_6$ panel for clarity.

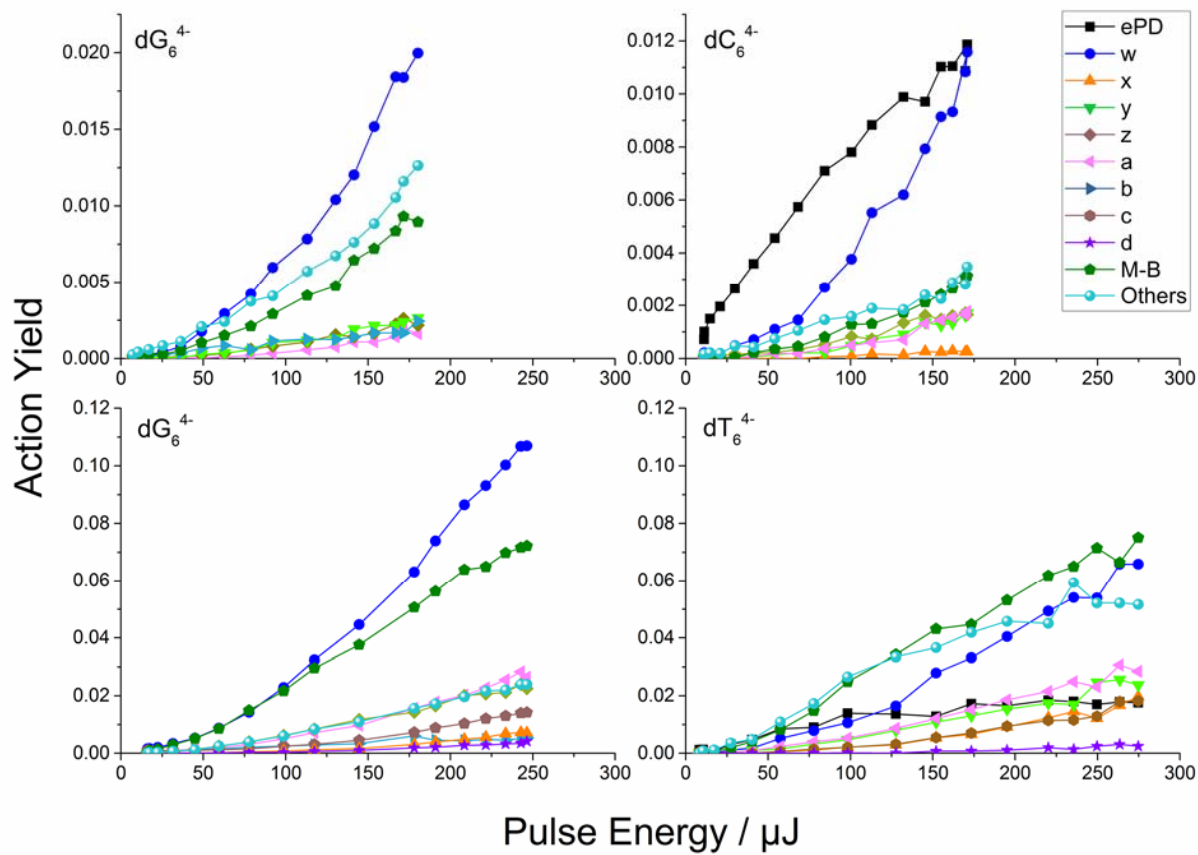

Figure S20. Action yield versus pulse energy for fragments assigned to different families for the 4- charge states of $G_6$ (top left), $C_6$ (top right), $A_6$ (bottom left) and $T_6$ (top right). Note the choice of colour reflects the colours used in the annotated MS. None of these are linear, and no individual component is either. The ePD is omitted from the $G_6$ panel for clarity.

| dG$_6^{2-}$ | P1 | P2 | P3 | P4 | P5 |
|---|---|---|---|---|---|
| cf1 | -0.33 | -0.33 | -0.33 | -0.5 | -0.5 |
| cf2 | -0.5 | -0.5 | 0 | -1 | 0 |
| cf3 | -0.5 | -0.5 | 0 | -0.5 | -0.5 |
| cf4 | -0.33 | -0.33 | -0.33 | -0.5 | -0.5 |

| dG$_6^{3-}$ | P1 | P2 | P3 | P4 | P5 |
|---|---|---|---|---|---|
| cf1 | -1 | -0.5 | -0.5 | -0.5 | -0.5 |
| cf2 | -0.5 | -0.5 | -0.5 | -0.5 | -1 |
| cf3 | -1 | -0.33 | -0.33 | -0.33 | -1 |
| cf4 | -1 | -0.33 | -0.33 | -0.33 | -1 |
| cf5 | -0.5 | -0.5 | -0.5 | -0.5 | -1 |
| cf6 | -1 | -0.33 | -0.33 | -0.33 | -1 |

| dG$_6^{4-}$ | P1 | P2 | P3 | P4 | P5 |
|---|---|---|---|---|---|
| cf1 | -1 | -1 | 0 | -1 | -1 |
| cf2 | 0 | -1 | -1 | -1 | -1 |

| dA$_6^{2-}$ | P1 | P2 | P3 | P4 | P5 |
|---|---|---|---|---|---|
| cf1 | -0.33 | -0.33 | -0.33 | -0.5 | -0.5 |
| cf2 | -0.33 | -0.33 | -0.5 | -0.5 | -0.33 |
| cf3 | -0.33 | -0.33 | -0.33 | -0.5 | -0.5 |

| dA$_6^{3-}$ | P1 | P2 | P3 | P4 | P5 |
|---|---|---|---|---|---|
| cf1 | -1 | -0.5 | -0.5 | -0.5 | -0.5 |
| cf2 | -1 | -0.33 | -0.33 | -0.33 | -1 |
| cf3 | -1 | -0.5 | -0.5 | -0.5 | -1 |
| cf4 | -1 | -0.33 | -0.33 | -0.33 | -1 |
| cf5 | -1 | -0.5 | -0.5 | -0.5 | -0.5 |

| dC$_6^{2-}$ | P1 | P2 | P3 | P4 | P5 |
|---|---|---|---|---|---|
| cf1 | -0.33 | -0.33 | -0.33 | -0.5 | -0.5 |
| cf2 | -0.33 | -0.33 | -0.33 | -0.5 | -0.5 |

| dC$_6^{3-}$ | P1 | P2 | P3 | P4 | P5 |
|---|---|---|---|---|---|
| cf1 | -1 | -0.33 | -0.33 | -0.33 | -1 |
| cf2 | -1 | -0.33 | -0.33 | -0.33 | -1 |
| cf3 | -1 | -0.33 | -0.33 | -0.33 | -1 |
| cf4 | -1 | -0.5 | -0.5 | -0.5 | 0.5 |
| cf5 | -1 | -0.33 | -0.33 | -0.33 | -1 |
| cf6 | -1 | -0.33 | -0.33 | -0.33 | -1 |

| dC$_6^{4-}$ | P1 | P2 | P3 | P4 | P5 |
|---|---|---|---|---|---|
| cf1 | -1 | -1 | -0.5 | -0.5 | -1 |

| dT$_6^{2-}$ | P1 | P2 | P3 | P4 | P5 |
|---|---|---|---|---|---|
| cf1 | -0.33 | -0.33 | -0.33 | -0.5 | -0.5 |
| cf2 | -0.33 | -0.33 | -0.5 | -0.5 | -0.5 |
| cf3 | -0.33 | -0.33 | -0.33 | -0.5 | -0.5 |
| cf4 | -0.5 | -0.5 | -0.33 | -0.33 | -0.33 |

| dT$_6^{3-}$ | P1 | P2 | P3 | P4 | P5 |
|---|---|---|---|---|---|
| cf1 | -0.5 | -0.5 | -0.5 | -0.5 | -1 |
| cf2 | -0.5 | -0.5 | -0.5 | -0.5 | -1 |
| cf3 | -1 | -0.33 | -0.33 | -0.33 | -1 |
| cf4 | -0.5 | -0.5 | -0.5 | -0.5 | -1 |

Figure S21. Summary of charge localization (linked to proton distribution in proximity to each of the phosphate groups) in each optimized "conformer" (dX$_6^{4-}$: 1 proton; dX$_6^{3-}$: 2 protons; dX$_6^{2-}$: 3 protons). We take the crude assumption that a phosphate group bear a charge = -1 and a neutralized phosphate group by a proton has a charge = 0. When a proton is shared between two phosphate groups, the charge is -0.5, and when 2 proton are shared between three phosphate groups, the charge of -0.33. The DFT calculations show that the charges are much more spread on the structure. The table is only helping to infer next to which phosphate group(s) the protons are located. The colors mean the following. When one proton is shared between two phosphate groups, the phosphate charge is assigned (-0.5 -0.5), written in the same color. When two protons are shared between 3 phosphates groups, they are assigned respective charges of (-0.33, -0.33, -0.33), with the same color. A phosphate group neutralized by a proton not shared by other groups is marked in green. This occurs only for some conformers of dG$_6$. Note that the phosphate groups with value of -1 (not neutralized) are all nevertheless making H-bonds with base or sugar groups, a phenomenon that "dilutes" the charges in practice.

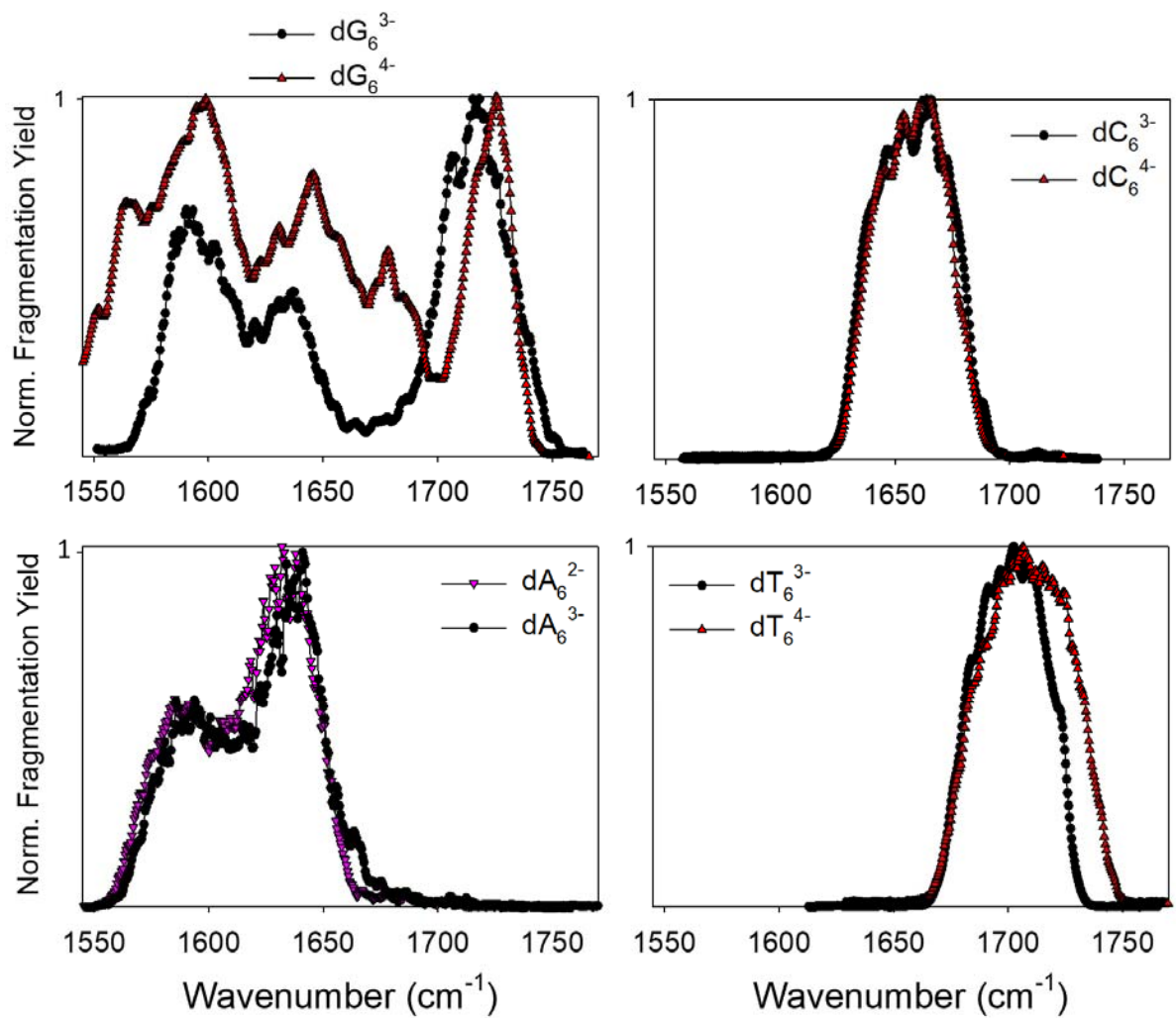

Figure S22. Experimental IRMPD spectrum of the single strands, as annotated in the legends. For the 3-single strands, the fragments (according to Mc Luckey nomenclature) taken into account are $d_1$-$H_2O^-$, $w_1^-$, $w_4^{2-}$, $w_2^{2-}$, $a_5$-$B^{2-}$, $a_3$-$B^-$, $y_3^-$, $w_3^-$, $a_4$-$B^-$, and base loss from the strand ss-$B^{2-}$ and internal fragments. For the other charge states, the fragments are of the same nature.

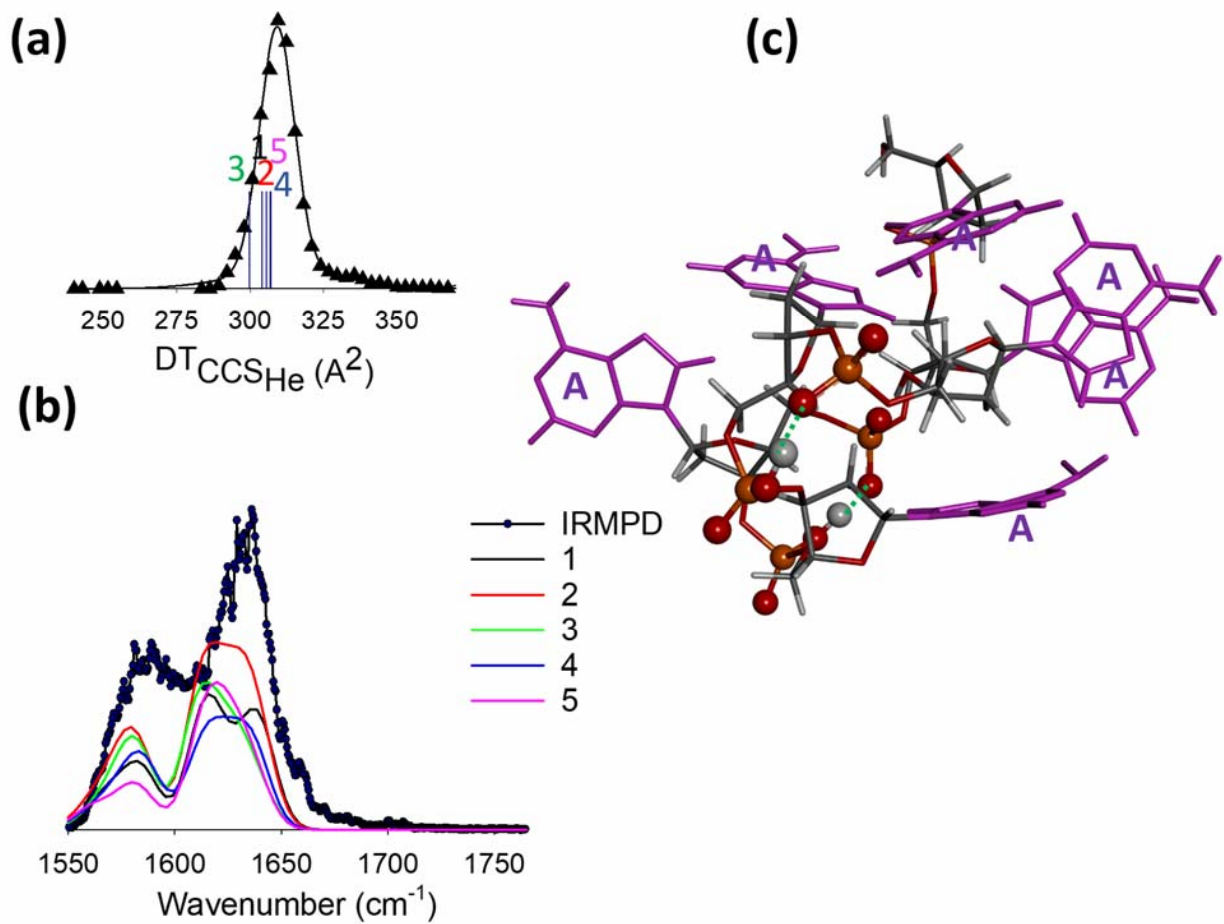

Figure S23. A) Ion mobility spectrometry of the single strand dA$_6^{3-}$ ($^{DT}$CCS$_{He}$). The vertical bars are the theoretical collisional cross section obtained on the different conformers of dA$_6^{3-}$ using the trajectory model (TM, mobcal). B) Experimental IRMPD spectrum of the single strand dA$_6^{3-}$ and the calculated IR spectra of different conformers of dA$_6^{3-}$. The calculations were performed using Gaussian 16 rev. B01 (DFT, B3LYP, 6-31G(d,p)+GD3) and frequencies were scaled by a factor of 0.97. C) Structure of conformer 5 of dA$_6^{3-}$. Hydrogen bonds between two pairs of phosphate groups are represented in ball and stick.

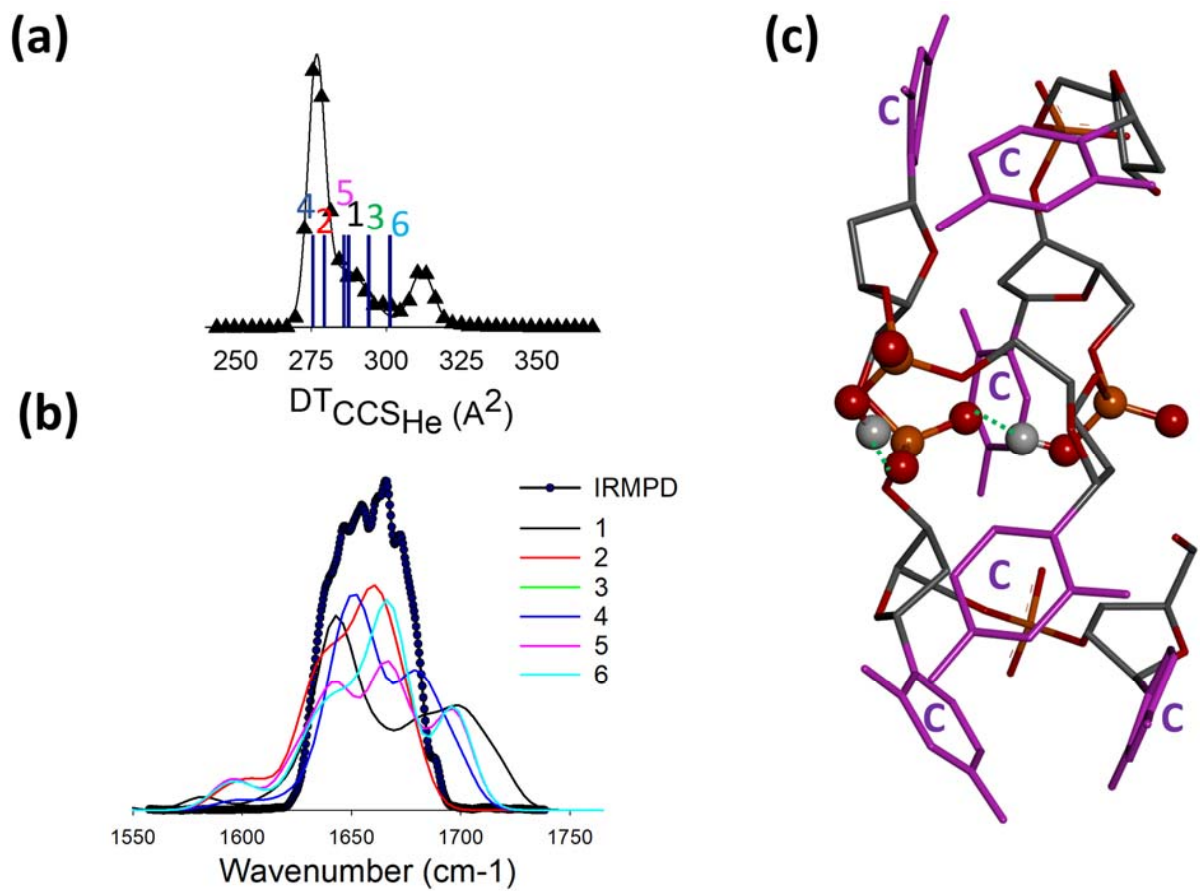

Figure S24. A) Ion mobility spectrometry of the single strand dC$_6$$^{3-}$ ($^{DT}$CCS$_{He}$). The vertical bars are the theoretical collisional cross section obtained on the different conformers of dC$_6$$^{3-}$ using the trajectory model (TM, mobcal). B) Experimental IRMPD spectrum of the single strand dC$_6$$^{3-}$ and the calculated IR spectra of different conformers of dC$_6$$^{3-}$. The calculations were performed using Gaussian 16 rev. B01 (DFT, B3LYP, 6-31G(d,p)+GD3) and frequencies were scaled by a factor of 0.97. C) Structure of conformer 5 of dC$_6$$^{3-}$. Hydrogen bonds between two pairs of phosphate groups are represented in ball and stick.

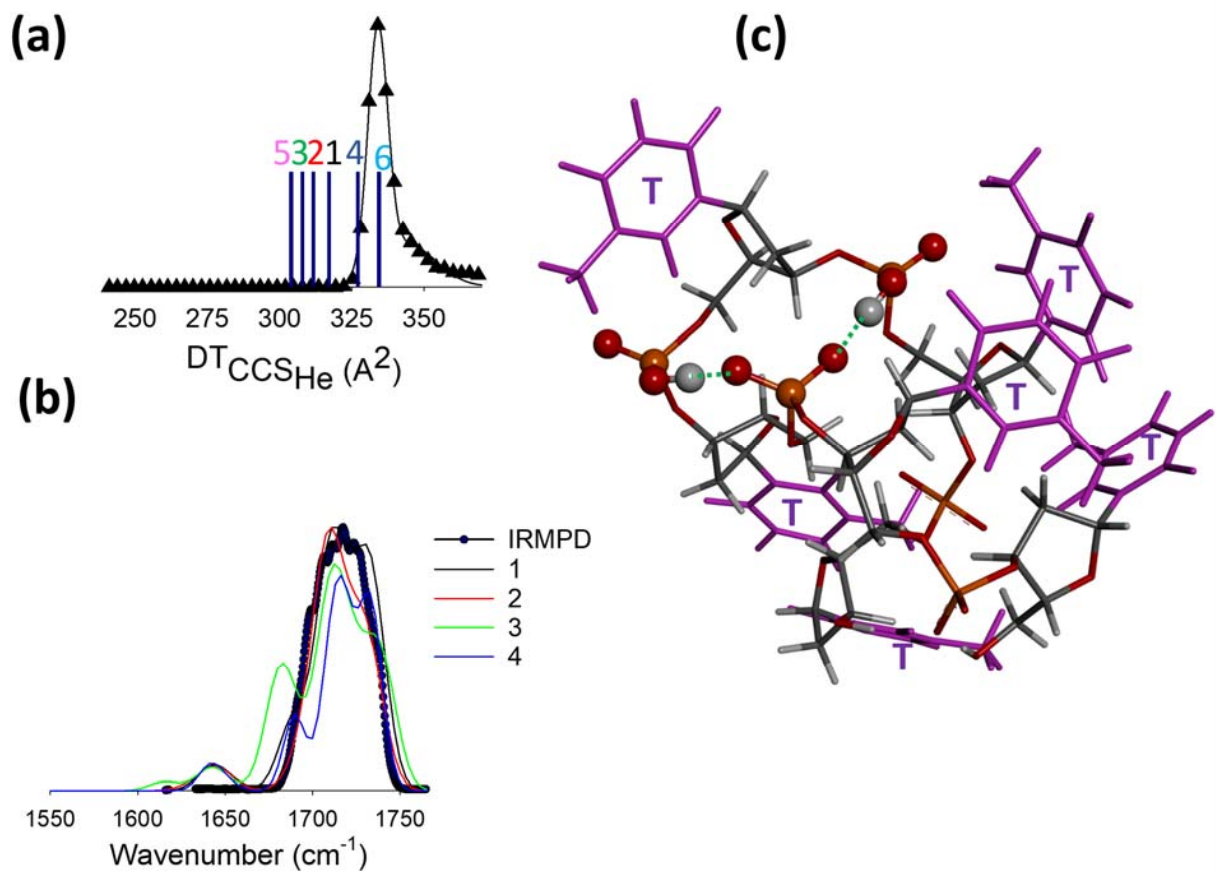

Figure S25. A) Ion mobility spectrometry of the single strand $dT_6^{3-}$ ($^{DT}CCS_{He}$). The vertical bars are the theoretical collisional cross section obtained on the different conformers of $dT_6^{3-}$ using the trajectory model (TM, mobcal). B) Experimental IRMPD spectrum of the single strand $dT_6^{3-}$ and the calculated IR spectra of different conformers of $dT_6^{3-}$. The calculations were performed using Gaussian 16 rev. B01 (DFT, B3LYP, 6-31G(d,p)+GD3) and frequencies were scaled by a factor of 0.97. C) Structure of conformer 3 of $dT_6^{3-}$. Hydrogen bonds between two pairs of phosphate groups are represented in ball and stick.

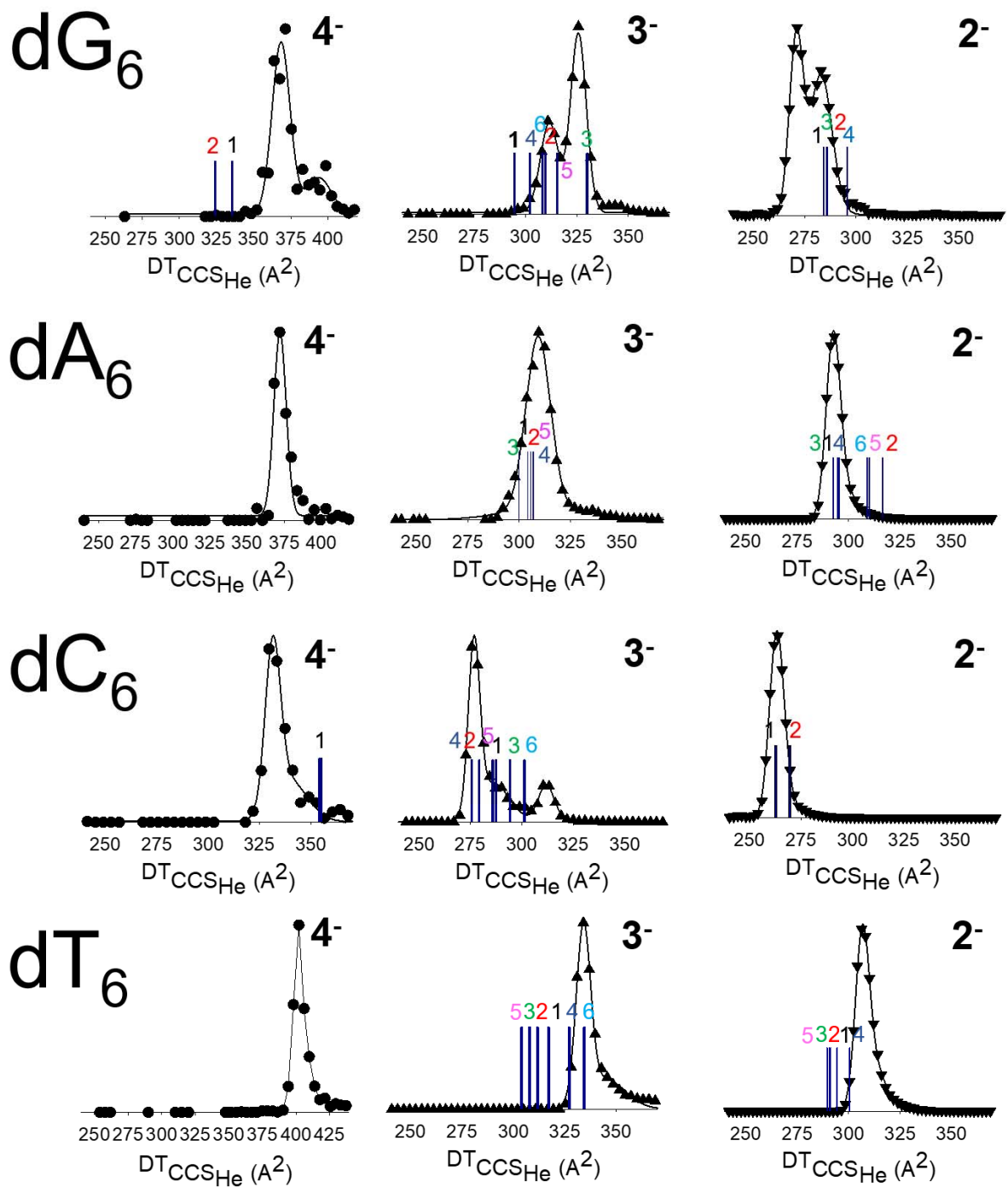

Figure S26. Collision cross section distributions ($^{DT}CCS_{He}$) for all charge states, and matching of the calculated collision cross section distributions (TM calculation, Mobcal, helium, 300K) for each conformer.

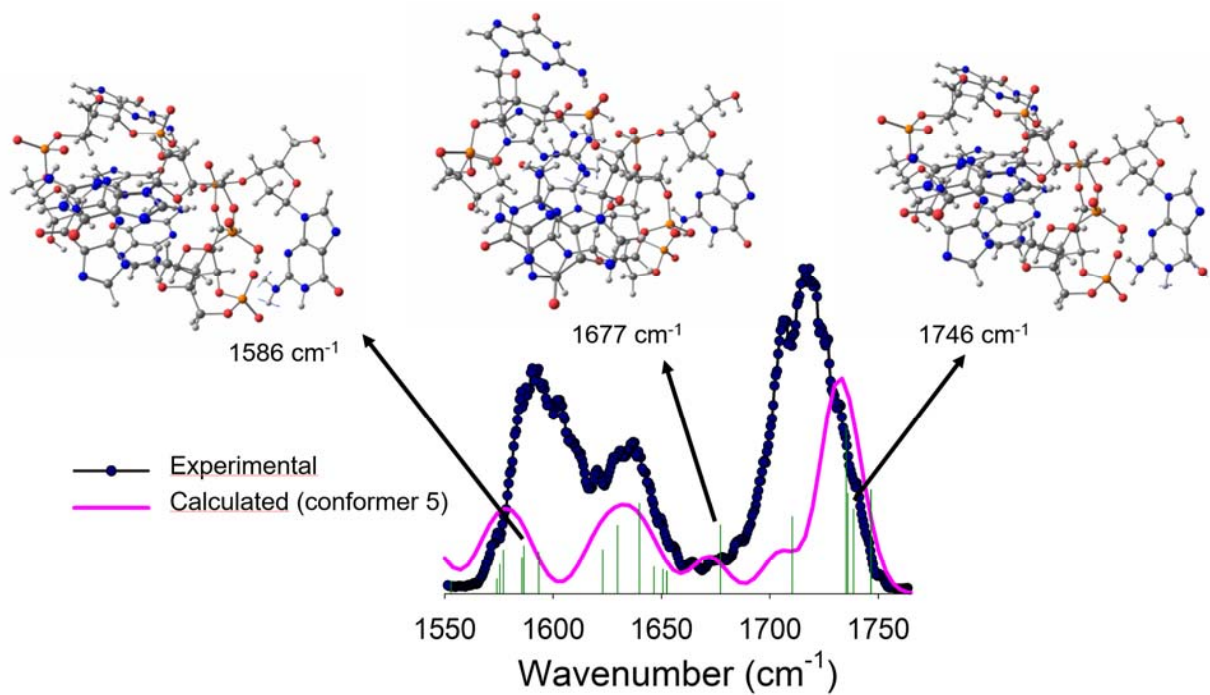

Figure S27. Examples of vibrational modes in the calculated spectra. An supporting animated version of this figure is available in PowerPoint format (FigureS23.ppt). The animation is visible by activating the "presentation" mode.

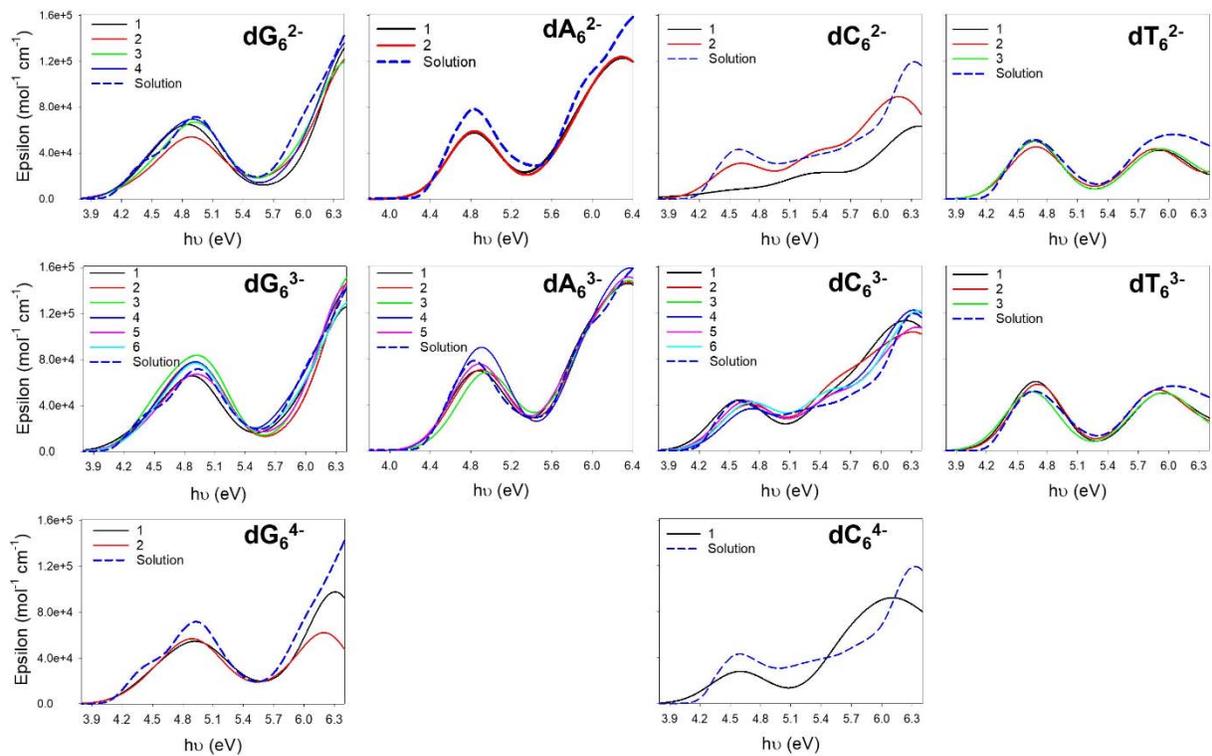

Figure S28. Comparison of experimental solution absorption spectra (dashed blue line) and calculated gas-phase absorption spectra (TD-DFT/MO6-2X/6-31G(d,p)) for all calculated conformers (the legend refers to the conformer number).

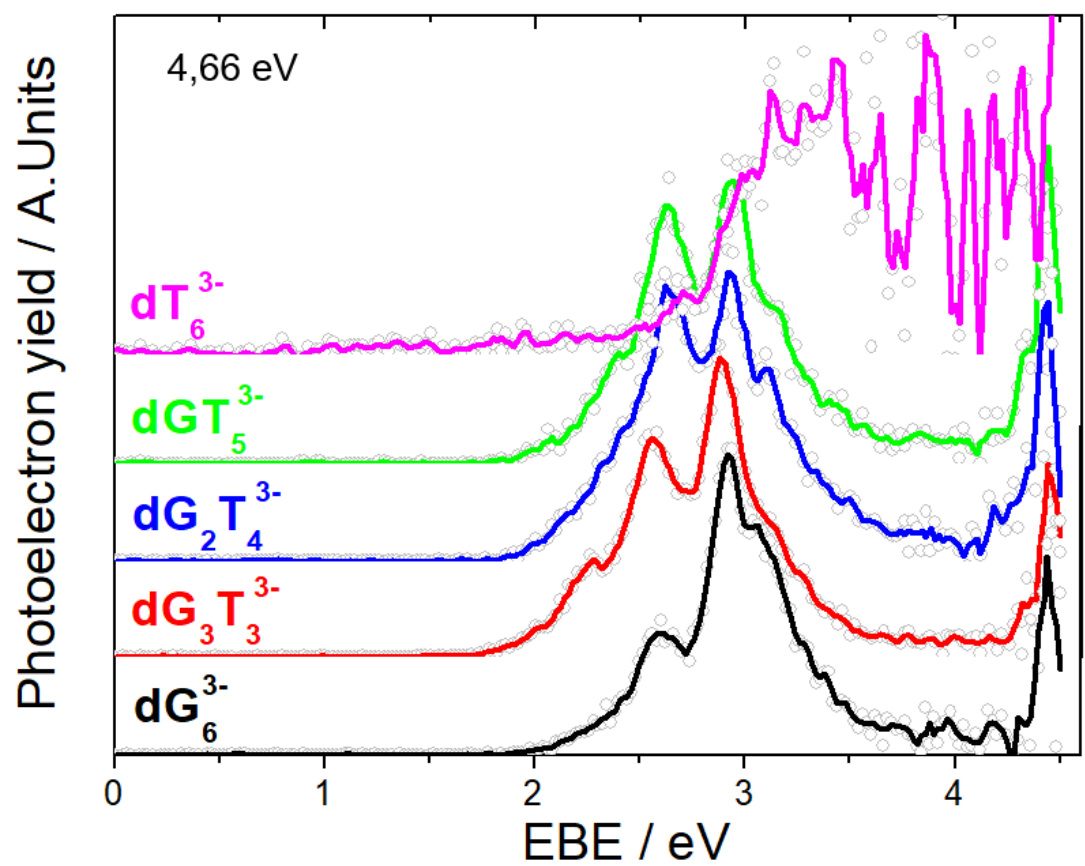

Figure S28. Photoelectron spectra of $dT_6^{3-}$, $dGT_5^{3-}$, $dG_2T_4^{3-}$, $dG_3T_3^{3-}$, and $dG_6^{3-}$, with an excitation wavelength of 266 nm ($h\nu$ = 4.66 eV).

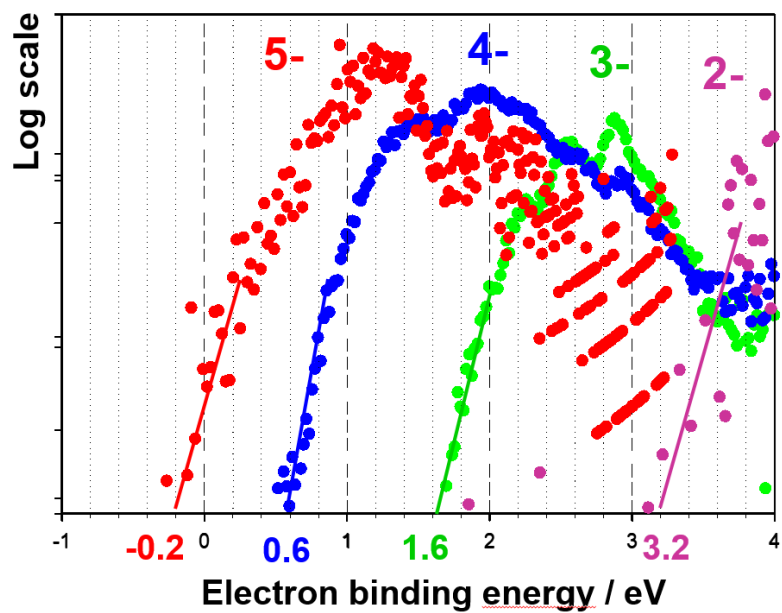

Figure S30. Photoelectron spectra of dG$_3$T$_3$ at charge states 5-, 4-, 3- and 2- (presented on a log scale to estimate the adiabatic detachment energy), recorded with an excitation wavelength of 266 nm (hν = 4.66 eV).

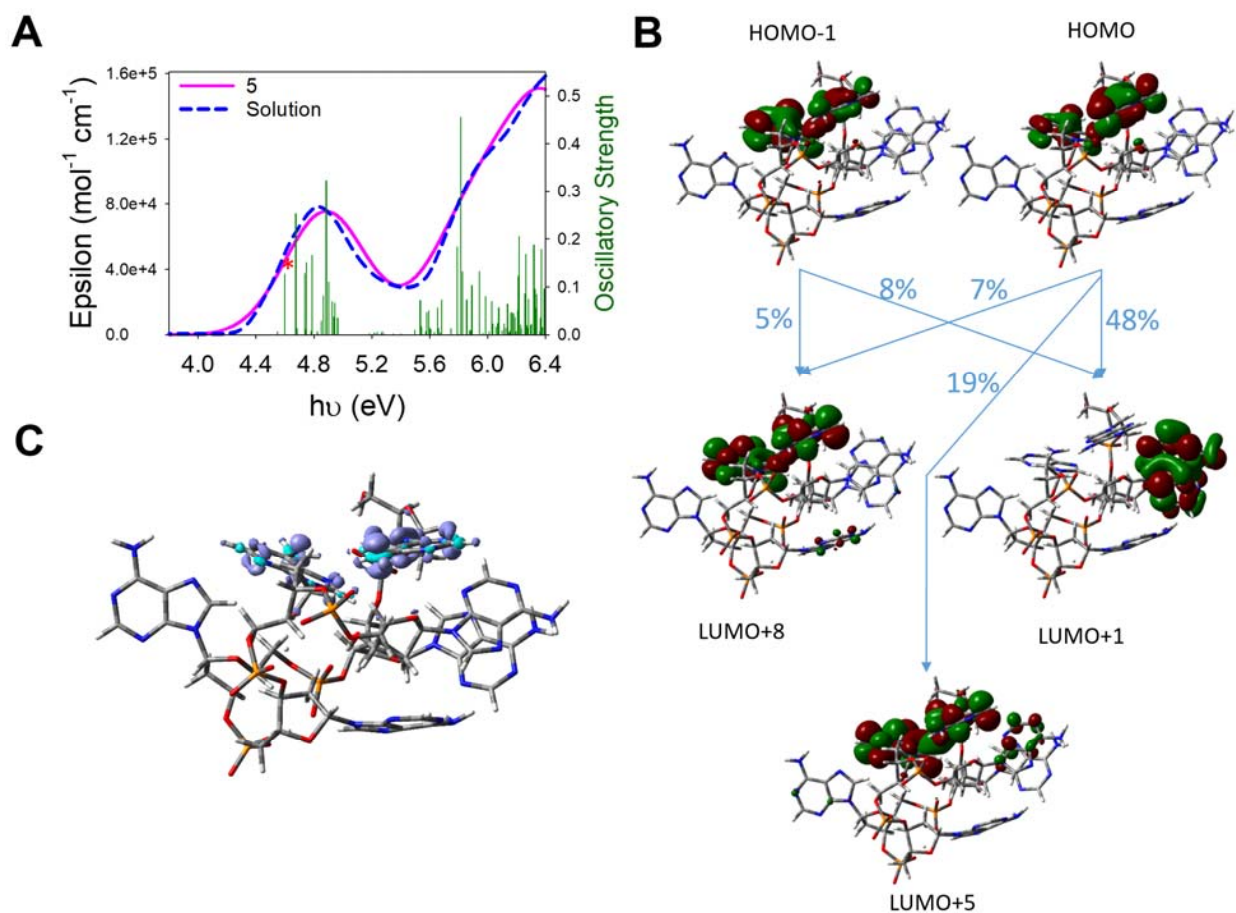

Figure S31. A) Experimental solution absorption spectrum and calculated gas-phase absorption spectrum (conformer 5) of dA$_6^{3-}$. B) Molecular orbitals involved in the first electronic transition of significant oscillator strength (shown by a star in panel A). C) Calculated difference of electronic densities between dA$_6^{3-}$ and the product of vertical electron detachment, dA$_6^{2-•}$.

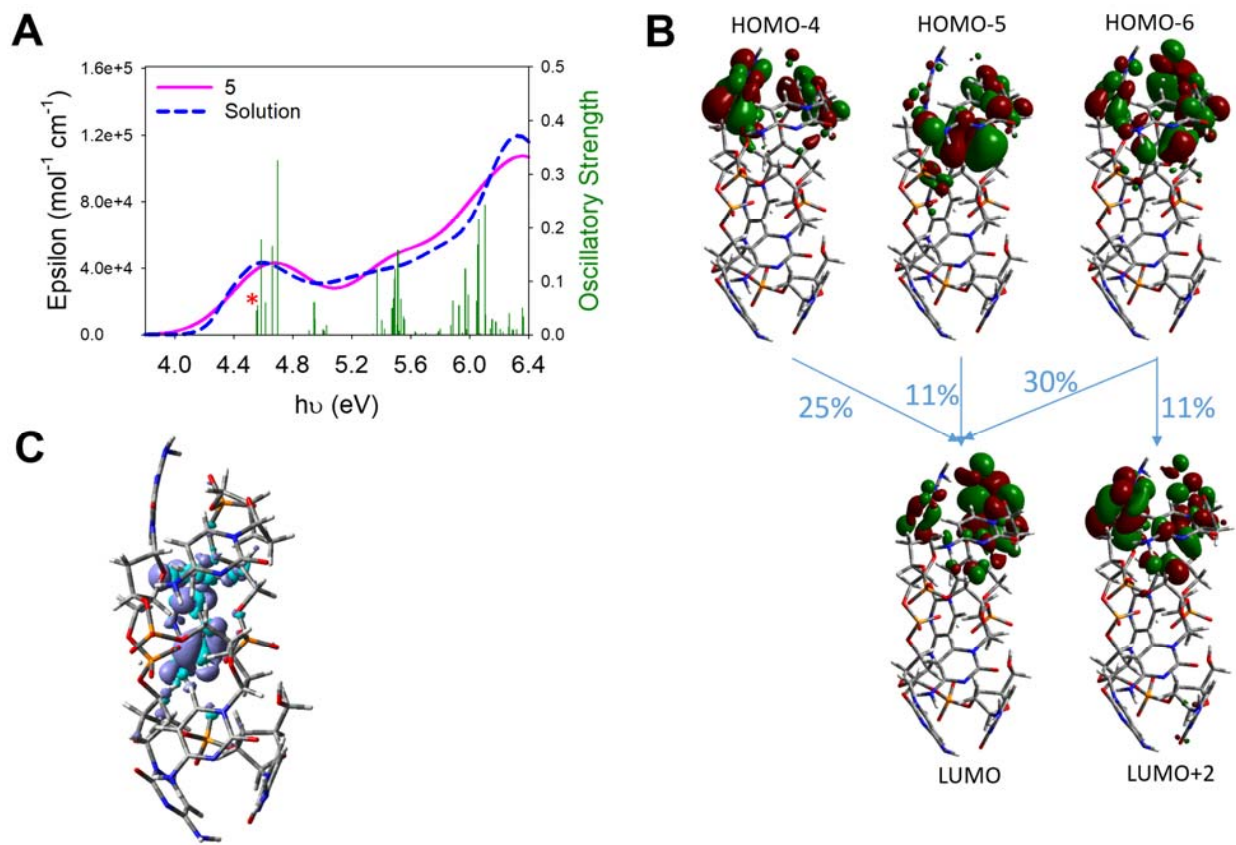

Figure S32. A) Experimental solution absorption spectrum and calculated gas-phase absorption spectrum (conformer 5) of dC$_6^{3-}$. B) Molecular orbitals involved in the first electronic transition of significant oscillator strength (shown by a star in panel A). C) Calculated difference of electronic densities between dC$_6^{3-}$ and the product of vertical electron detachment, dC$_6^{2-\bullet}$.

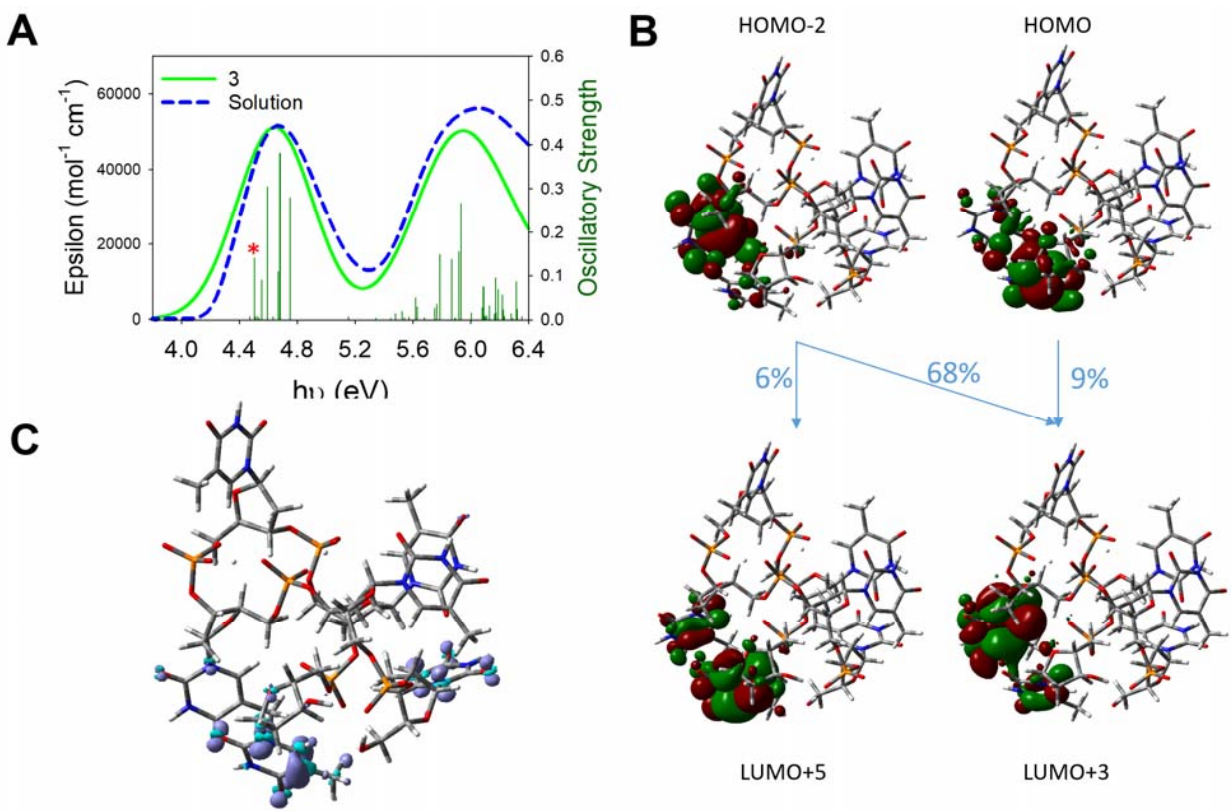

Figure S33. A) Experimental solution absorption spectrum and calculated gas-phase absorption spectrum (conformer 3) of dT$_6^{3-}$. B) Molecular orbitals involved in the first electronic transition of significant oscillator strength (shown by a star in panel A). C) Calculated difference of electronic densities between dT$_6^{3-}$ and the product of vertical electron detachment, dT$_6^{2-\bullet}$.